\documentclass[11pt]{article}
\usepackage[myheadings]{fullpage}
\usepackage{setspace}
\usepackage{amsmath}
\usepackage[dvips]{graphicx}
\usepackage[round,numbers,sort&compress]{natbib}
\usepackage{dcolumn}
\usepackage{bm}
\usepackage{latexsym}
\date{}

\begin{document}
\begin{titlepage} 
\thispagestyle{empty}

\title{
\textbf{Microtubule depolymerization by the kinesin-8 motor
     Kip3p: a mathematical model.}}

 \author{L. E. Hough$^{1}$\\
 Physics Department  \\
 \and
 Anne Schwabe$^{1}$\\
 Chemistry and Biochemistry Department \\
 \and
 Matthew A. Glaser\\
 Physics Department \\
 \and
 J. Richard McIntosh\\
 MCD Biology Department\\
 \and
 M. D. Betterton\\
 Physics Department\\
\\
 University of Colorado at Boulder, Boulder CO}
\footnotetext[1]{These authors contributed equally. \\
\parbox[t]{0.22in} \mbox{\textbf{Address reprint requests to:}} 
M. D.  Betterton \\
\parbox[t]{0.22in} \mbox{\textbf{Phone:}}  303-735-6235\\
\parbox[t]{0.22in} \mbox{\textbf{E-mail:}} mdb@colorado.edu }

\end{titlepage}
\maketitle
\newpage
\setcounter{page}{1}

\section*{\textrm{ABSTRACT}}
Proteins from the kinesin-8 family promote microtubule (MT)
depolymerization, a process thought to be important for the control of
microtubule length in living cells. In addition to this MT shortening
activity, kinesin 8s are motors that show plus-end directed motility
on MTs. Here we describe a simple model that incorporates directional
motion and destabilization of the MT plus end by kinesin 8.  Our model
quantitatively reproduces the key features of length-vs-time traces
for stabilized MTs in the presence of purified kinesin 8, including
length-dependent depolymerization.  Comparison of model predictions
with experiments suggests that kinesin 8 depolymerizes processively,
i.e., one motor can remove multiple tubulin dimers from a stabilized
MT.  Fluctuations in MT length as a function of time are related to
depolymerization processivity.  We have also determined the parameter
regime in which the rate of MT depolymerization is length dependent:
length-dependent depolymerization occurs only when MTs are
sufficiently short; this crossover is sensitive to the bulk motor
concentration.

\newpage

\section*{Introduction}

Regulation of microtubule (MT) length is an important cellular
process.  Abnormal MT lengths can mislocalize the nucleus or mitotic
spindle and cause defects in polarized cell growth or mitosis.
Although MT length regulation is not fully understood, several
mechanisms have been proposed, including chemical gradients in the
mitotic spindle \cite{CaudronScience2005}, cortical interactions
\cite{YehMBC2000}, trafficking of proteins that bind to MT ends
\cite{BielingNature2007}, and a balance between proteins that promote
MT polymerization and depolymerization
\cite{HowardCOCB2007,goshima05}.

Kinesin-8 proteins appear to help regulate MT length \textit{in vivo}.
Deletion of kinesin-8 genes leads to longer interphase and spindle MTs
and defects in mitosis \cite{WestMBC2001,WestJCS2002,
  MayrCB2007,GarciaCB2002, GarciaEMBO2002,BusterMBC2007,GuptaNCB2006},
suggesting that kinesin 8s promote MT depolymerization. In addition,
kinesin-8 motors show processive, plus-end directed motility on MTs
\textit{in vivo} \cite{GuptaNCB2006,stumpff08}.  Recent work has shown
that kinesin-8 proteins are important in chromosome oscillations and
MT length fluctuations \cite{WestJCS2002,stumpff08,unsworth08}.

Biochemical experiments with some purified kinesin 8s have reproduced
many of observations made \textit{in vivo}: kinesin 8 moves
processively toward the MT plus end \cite{VargaNCB2006} and the MTs
then depolymerize, even when stabilized with taxol or GMPCPP
\cite{VargaNCB2006, GuptaNCB2006, MayrCB2007}. Kinesin 8s are thus
thought to be able to remove the GTP-tubulin cap that stabilizes
growing MTs \textit{in vivo}.  Varga et al.~recently proposed that the
processive motility of kinesin 8s preferentially increases their
concentration at the ends of longer MTs \cite{VargaNCB2006}, thereby
more rapidly depolymerizing longer MTs.  They proposed that this
length-dependent activity serves to regulate MT length.  Howard and
Hyman proposed that a constant MT growth rate coupled with
length-dependent depolymerization would result in a tighter
distribution of MT lengths than that set by dynamic instability
\cite{HowardCOCB2007}. In particular, MT dynamic instability gives an
exponential distribution of MT lengths
\cite{MitchisonNature1984b,bicout97}, while a coupling between MT
growth and length-dependent depolymerization could give a tighter
distribution \cite{govindan08}. The \textit{in vitro} experiments are
consistent with the \textit{in vivo} observation of longer MTs when
the kinesin 8 is deleted, but the reasons why kinesin 8 depletion and
overexpression alter mitotic oscillations are less clear.

Given that not all of the \textit{in vivo} results can be simply
understood in terms of the \textit{in vitro} observations, we sought
to determine if a detailed theory of the \textit{in vitro} experiments
could lead to insights into the behavior of kinesin 8s in cells.  We
have developed a simple mathematical model that quantitatively
captures the length-vs-time traces of stabilized MTs in the presence
of purified kinesin 8.  The results are consistent with processive MT
depolymerization by kinesin 8, i.e., multiple tubulin dimers can be
sequentially removed by a single motor.  We explored a consequence of
this processive depolymerization: altered fluctuations in MT length
during depolymerization.  In addition, we studied the distribution of
motors along the MT and find that length-dependent MT depolymerization
occurs only for sufficiently short MTs, i.e., those below a
``crossover'' length, which depends strongly on the bulk motor
concentration and model parameters.
 
Our work builds on previous physical theory that considered the motion
of multiple motors on a one dimensional track: Parmeggiani et
al.~studied a model similar to ours but disallowed changes in the
length of the track \cite{parmeggiani04}.  Nowak et al.~extended this
work to allow track lengthening catalyzed by motors \cite{nowak07}.
Other theoretical papers have focused specifically on MT
depolymerization by kinesins.  For example, the coupling between motor
motion and MT depolymerization is thought to be important for
kinesin-13 proteins, which also depolymerize GMPCPP- and
taxol-stabilized MTs \cite{HeleniusNature2006}. An important
difference from the present work is that the kinesin-13 protein MCAK
moves on MTs by diffusion along the MT lattice, not through motor
activity, and it accumulates at the MT ends through a binding
preference for this part of the MT.  A previously developed theory of
MT shortening by MCAK focused on the dynamic accumulation of motors at
the MT end \cite{KleinPRL2005}.  Both the experiments and theory on
MCAK found evidence for processive depolymerization.  Finally, a model
of MT shortening catalyzed by either kinesin-8 or -13 motors was
recently introduced \cite{govindan08}.

Our work extends previous theories of motor motion and
depolymerization in important ways. 
The mathematical model we consider is similar to that of Govindan et
al.~\cite{govindan08}. The key differences are that Govindan et
al.~neglected motor crowding effects and analyzed the steady state of
the model, while here we include crowding effects and non-steady-state
dynamics.
An important theoretical challenge arises because the rate at which
motors accumulate at the MT end and the rate of MT length change are
similar (as observed in experiments, see
refs.~\cite{VargaNCB2006,GuptaNCB2006}).  A steady-state analysis,
while mathematically more tractable, may therefore miss dynamic
behavior present in the experiments. In this paper we first consider a
kinetic Monte-Carlo simulation of the full model.  Then we develop a
mean-field model that explicitly considers non-steady-state effects by
describing the coupling between the dynamics of the MT end and the
motor occupancy at the end.  The depolymerization rates we obtain in
the mean-field model agree well with Monte-Carlo simulations of the
full model. 
We compare the results of the model, including non-steady-state
effects, to the experimental data obtained by Varga et
al.~\cite{VargaNCB2006} on length-dependent depolymerization of MTs by
Kip3p.

\section*{A mathematical model of kinesin-8 action on MTs
  quantitatively reproduces experiments}

\begin{figure}[t]
  \centering
  \includegraphics[width=18 cm]{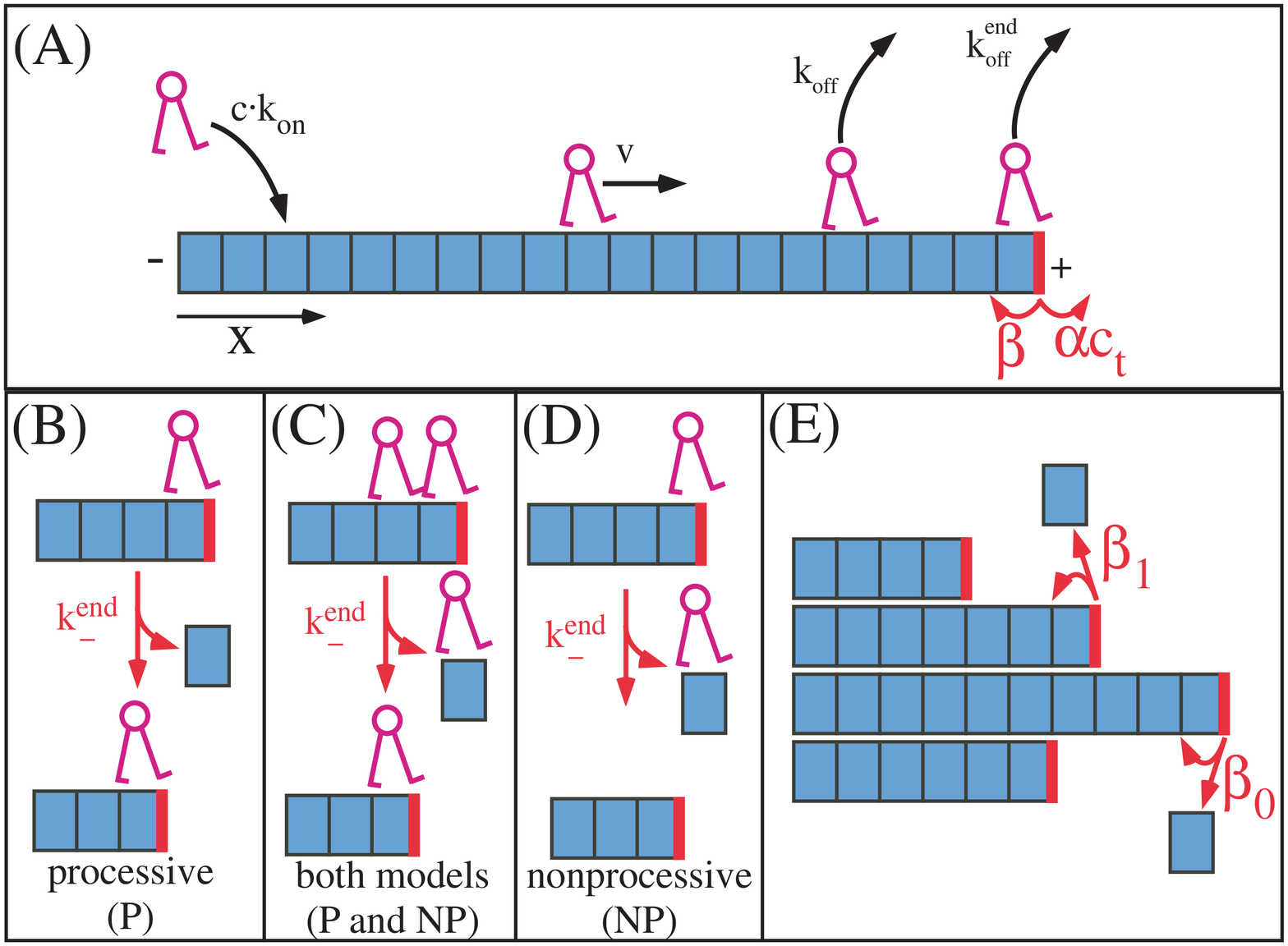}
  \caption {}
  \label{model}
\end{figure}

Our theoretical model of kinesin-8 motors includes biased motor motion
toward MT plus ends and motor-catalyzed depolymerization of MT plus
ends. A schematic of the model is shown in figure \ref{model}A.  The
position $x$, in units of tubulin dimers, is measured from the MT
minus end ($x=0$). We assume motors track individual protofilaments
and step from one tubulin dimer to the next (a length of 8 nm) toward
the MT plus end at rate $v$.  We assume the motors do not step
backward ($k_-=0$), as kinesin-8 motors have shown highly biased
motion \cite{VargaNCB2006}.

Motors can bind to and unbind from the MT. The ``on'' rate is
$k_{\mbox{\small on}} c$, where $c$ is the bulk concentration of
motors (typically assumed to be constant). The ``off'' rate is
$k_{\mbox{\small off}}$, unless the motor is bound at the end of the
MT.  Motors dissociate from the plus end of the MT at rate
$k_{\mbox{\small off}}^{\mbox{\small end}}$. We neglect any special
affinity that the motors may have for the MT minus end, since their
plus-end directed motility makes their occupancy at the minus end
negligible. We also assume that motors have a negligible affinity for
soluble tubulin dimers. Although the binding affinity of kinesin-8
motors for soluble tubulin is currently unknown \cite{GuptaNCB2006},
when we allowed motor binding to free tubulin with an affinity as
large as the affinity for MT dimers, there was little change in the
results (see below, and other data not shown). Therefore, the precise
value of the affinity is not important for our results.

We considered two distinct depolymerization mechanisms: one assumes
processive depolymerization (P) and another assumes purely
non-processive depolymerization (NP).  In the first case (P) motors
processively depolymerize protofilaments by removing one dimer and
stepping backward at rate $k_-^{\mbox{\small end}}$.  If there is
another motor bound behind the depolymerizing motor, the motor at the
end is assumed to unbind. This assumption is motivated by live-cell
imaging of fluorescent Kip3p, which showed that a clump of motors
accumulates at the MT plus end during MT growth, but the clump
fluorescence greatly decreases during MT shrinkage
\cite{VargaNCB2006,GuptaNCB2006}.  Since protofilaments are straight
in the bulk of the MT and protofilament curvature is associated with
depolymerization \cite{muller-reichert98}, our picture of processive
depolymerization is consistent with a structural mechanism in which
the motor has a higher affinity for a straight protofilament than a
curved region of the protofilament.  Our picture of processive
depolymerization is also consistent with individual motors inducing
catastrophe (in dynamic MTs) and thereby removing large numbers of
tubulin dimers.  In the absence of crowding effects (if motors fall
off before reaching another motor), then the number of dimers removed
per motor is $a=k_-^{\mbox{\small end}}/k_{\mbox{\small
    off}}^{\mbox{\small end}}$.  Since motor crowding reduces the
depolymerization processivity, the value $a$ will
only be observed for sufficiently low motor concentration.  Note
that this picture of depolymerization can apply either to a motor that
directly catalyzes depolymerization, or a motor that recruits to the
MT end another protein that promotes depolymerization.

In the second depolymerization mechanism (NP) motors are assumed to
remove a single dimer and fall off with that dimer.  In this case,
each motor removes at most one dimer (a motor could unbind before
removing a dimer).

Our idealized MTs consist of 13 protofilaments arranged cylindrically
(we neglect helical arrangement of protofilaments).  In all modeling,
we further assume that dimer removal by motors is non-cooperative,
both within and between protofilaments.  If each motor acts
independently (the case of non-cooperative motors), then the rate of
depolymerization is, on average, proportional to the density of motors
at the end of the MT. (Note that we do consider the possibility of
protofilament interactions in tubulin depolymerization; see below.)

Typically we consider motor-induced depolymerization of stable MTs, so
that the MT has no intrinsic dynamics. However, in some cases we use a
simple model of MT dynamics: dimers add to a MT end at rate $\alpha
c_t$ (where $c_t$ is the bulk concentration of tubulin dimers) and
dissociate at rate $\beta$ (fig.~\ref{model}). In one set of
simulations we allowed the depolymerization rate $\beta$ to depend on
the number of lateral interactions between protofilaments
\cite{vanburen02,vanburen05}.  In this case, the rate at which a
terminal tubulin dimer unbinds from a protofilament is given by
$\beta_0$ if the dimer has no lateral neighbors, $\beta_1$ if the
dimer has 1 lateral neighbor, and $\beta_2$ if the dimer has 2 lateral
neighbors.

We developed a kinetic Monte Carlo simulation of the full model, and
studied analytic approximations to the model.

\subsection*{Depolymerization of stabilized MTs}

In the minimal model of stabilized MTs (figure 1A-D), there are 
5 independent parameters. Three parameters can be derived directly
from the data of Varga \textit{et al.}~on the budding yeast kinesin-8
motor Kip3p \cite{VargaNCB2006}. (See table 1 for a summary of
estimated parameter values.) For these estimates, we assumed that each
motor and protofilament behaves independently: we neglected effects of
protofilament interactions, motor crowding, and motor depolymerization
cooperativity.  The measured motor velocity was 3.6 $\mu$m min$^{-1}$
which gives $v = 7.5$ dimers s$^{-1}$.  The average run length of a
motor was 12 $\mu$m, which gives $\ell = 1500$ dimers.  (Note that this
run length is likely a lower bound, since the experiments ignored
motors that reach the MT end when determining the run length.) Since
$\ell = v/k_{\mbox{\small off}}$, this implies $k_{\mbox{\small off}}=
5 \times 10^{-3}$ s$^{-1}$.  The typical residence time of a motor at
the end of the microtubule was half a minute, so $k_{\mbox{\small
    off}}^{\mbox{\small end}} = 3.3 \times 10^{-2}$ s$^{-1}$.

The two parameters $k_-^{\mbox{\small end}}$ and $k_{\mbox{\small
    on}}$ were not directly measured for Kip3p, but we can estimate
their values from the experimental data.  The maximum depolymerization
velocity observed was approximately 2 $\mu$m min$^{-1}$.  This
suggests $k_-^{\mbox{\small end}} \ge 4$ dimers s$^{-1}$. However, by
comparing our simulations with the data of Varga et al.~we found a
better fit using the value $k_-^{\mbox{\small end}} = 13$ dimers
s$^{-1}$ (see below).  We estimated the motor binding rate from the
kymograph of figure 3d from Varga et al., which shows 9 binding events
in 6 minutes on a MT 12 $\mu$m in length, for an unspecified bulk
motor concentration (of order 1 nM).  This gives a microscopic rate
constant of order $10^{-6}$ site$^{-1}$ nM$^{-1}$ s$^{-1}$.  We
determined the value $k_{\mbox{\small on}} = 3 \times 10^{-6}$
site$^{-1}$ nM$^{-1}$ s$^{-1}$ by comparing simulations to the
experimental data. In our model, the parameter with the greatest
uncertainty is $k_{\mbox{\small on}}$. We discuss most results using
our best-fit value and discuss below the consequences of varying this
value.

We first simulated intrinsically stable MTs.  The simulations started
with bare MTs (no motors bound), and motor binding was begun at $t=0$.
Using the values of $k_-^{\mbox{\small end}} = 13$ s$^{-1}$ and
$k_{\mbox{\small on}}= 3 \times 10^{-6}$ site$^{-1}$ nM$^{-1}$
s$^{-1}$, we found good qualitative agreement with experiments.  The
depolymerization rate in the simulations increased as the motor
occupancy along the MT increased.  The time required for equilibration
of the motor occupancy was approximately the typical time a motor
stayed bound, $1/k_{\mbox{\small off}}$ (see further discussion of
occupancy equilibration below in the section on motor occupancy
profiles).  The depolymerization rate later decreased once the MT
shortened sufficiently.

Our simulations of stabilized MTs did not exhibit the long-time
behavior observed in the Varga et al.~experiments \cite{VargaNCB2006}.
In these experiments, kinesin-8 motors were not able to fully
depolymerize GMPCPP-stabilized MTs: the depolymerization rate slowed
and dropped to zero over several minutes, leading to a ``long-time
tail'' in the plot of MT length as a function of time. This could have
occurred for several reasons: ({\it i}) the motor activity may have
decreased, e.g., due to depletion of ATP; ({\it ii}) motors may have
bound to free tubulin dimers or the coverglass, effects which would
deplete the concentration of free motors
\cite{VargaNCB2006,GuptaNCB2006}; ({\it iii}) motors may depolymerize
cooperatively, so that the decrease in motor concentration at the end
of a MT due to shortening had a nonlinear effect on the depolymerization
velocity; or ({\it iv}) the slow polymerization activity of GMPCPP
tubulin \cite{pmid1421572} may have prevented full MT
depolymerization, because tubulin dimers could add to the end of the
MT, increasing its length.

We found that motor binding to free tubulin is not likely to explain
the long-time tails. We simulated binding of motors to free tubulin dimers
with varying binding affinity. Even when the affinity is large---as
large as the affinity for tubulin in the MT lattice---we found only a
small change in the depolymerization dynamics (data not shown).

\begin{figure}[t]
  \centering
  \includegraphics[width=12 cm]{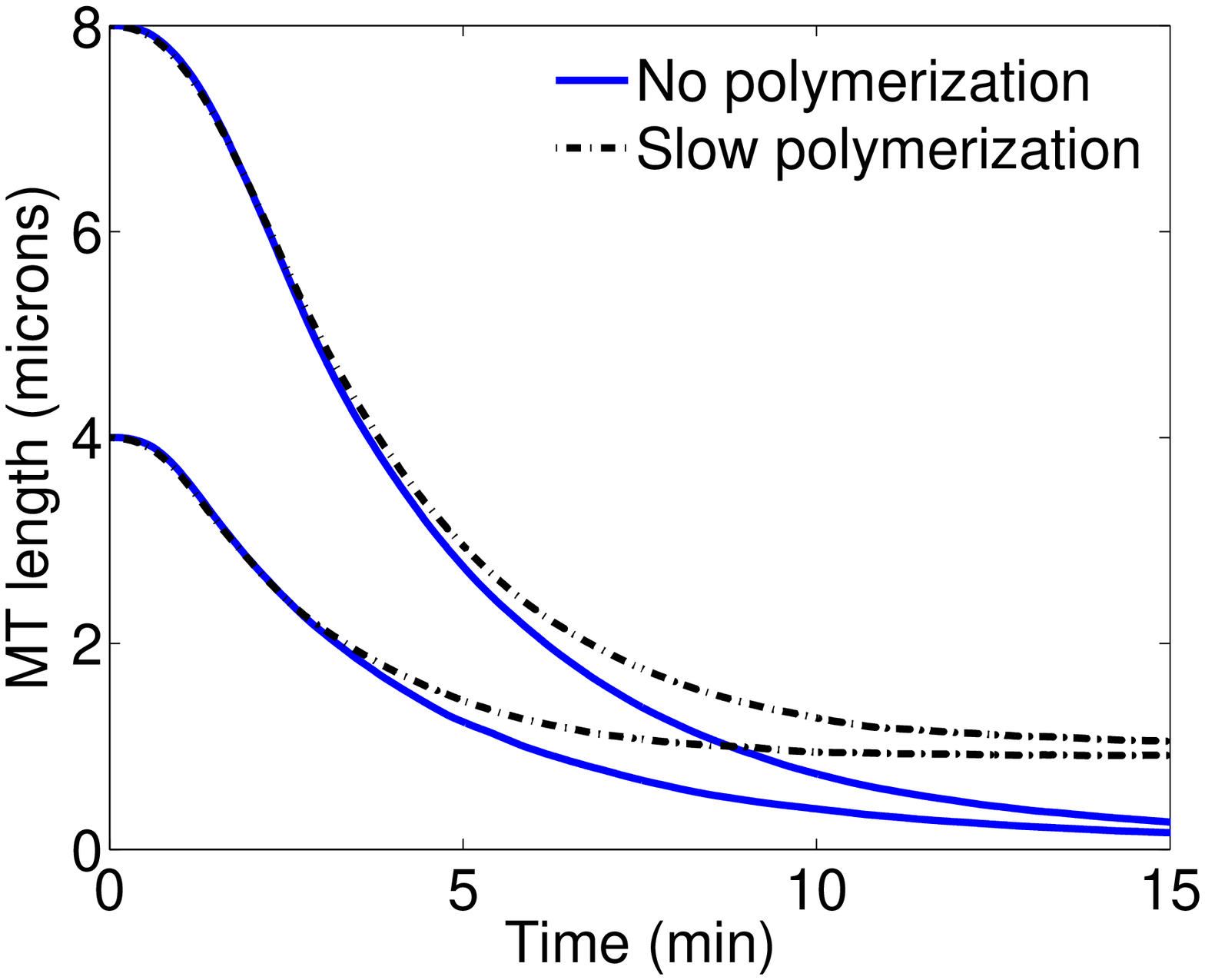}
  \caption {}
  \label{longtails}
\end{figure}

We then modeled polymerization and depolymerization of the MTs with
nonzero values of the rate constants $\alpha$ and $\beta$
(fig.~\ref{model}).  The intrinsic dynamics of the MT plus end could
have two effects on the motors at the MT end: (i) the motors could
remain attached to the tubulin dimer that is removed or (ii) the motor
could be displaced backward and remain on the MT.  We found that the
former led to significantly decreased MT depolymerization rates (the
intrinsic MT dynamics lead to the unbinding of many motors), while the
latter gave quantitative agreement with the \textit{in vitro}
experiments.  In particular, we estimated that the total tubulin
concentration in the Varga et al.~experiments is of order 100 nM.
Therefore, we included in the model the measured rates $\alpha = 5.4$
dimers $\mu$M$^{-1}$ s$^{-1}$ and $\beta = 0.1$ dimers s$^{-1}$
\cite{pmid1421572} and found quantitative agreement between the
simulation traces and the experiments (see figure \ref{longtails}).
In this case, the free tubulin concentration becomes high enough
that polymerization is large enough to balance motor-induced
depolymerization and the MTs approach a constant, time-independent
length.  Thus, we conclude that the accumulation of free tubulin
within the flow chambers, and the subsequent slow polymerization of
the MTs, can account for the long-time tails observed in the Varga et
al.~experiments \cite{VargaNCB2006}.

After showing that including the slow polymerization activity of
GMPCPP tubulin in the model is sufficient to resolve this qualitative
disagreement between experiments and theory, we fit the experimental
data to determine the unknown parameters.  With processive
depolymerization (model P), the only free parameters were
$k_-^{\mbox{\small end}}$, $k_{\mbox{\small on}}$, and $\beta_{0,1}$.
Figure \ref{variance} shows similar MT length versus time curves for
different values of these parameters.  In model P with independent
protofilaments ($\beta_{0,1}=0$) a lower depolymerization processivity
could (within some range) be offset by a higher on rate
(fig.~\ref{variance}A).  In addition, decreasing the stability of
neighborless protofilaments has similar effects to increasing the MT
depolymerization processivity (fig.~\ref{variance}C). With
nonprocessive depolymerization (model NP), only very high
$\beta_{0,1}$ give MT depolymerization time courses that match the
experimental results (see below, where nonprocessive depolymerization
is discussed in more detail).  The best-fit experimental traces were
obtained with 
$k_-^{\mbox{\small end}} = 13$ s$^{-1}$ and $k_{\mbox{\small on}}= 3
\times 10^{-6}$ site$^{-1}$ nM$^{-1}$ s$^{-1}$.

\section*{Evidence for processive depolymerization by kinesin-8
  motors} 

Processive depolymerization of MTs by motors is more consistent with
the parameters measured by Varga et al.~than non-processive
depolymerization. In particular, they found that an individual motor
is resident at the MT end for approximately 30 seconds (at low motor
density), and the depolymerization velocity they observed had a
maximum of 2 $\mu$m min$^{-1}$ (at higher motor density), suggesting
that (if residence times are similar over a range of motor densities)
approximately 125 tubulin dimers could be depolymerized during the
binding time of a single motor at the MT end \cite{VargaNCB2006}.  Our
best fits to the experimental data gave a higher maximum
depolymerization rate for individual motors.  We found the best fit
with $k_-^{\mbox{\small end}} = 13$ dimers s$^{-1}$.  This would imply
a maximum number of dimers removed per motor $a=k_-^{\mbox{\small
    end}}/k_{\mbox{\small off}}^{\mbox{\small end}}\sim 400$ dimers.
(Note that our simulations assumed no interactions between
protofilaments or motors, each of which could alter the apparent
number of dimers removed per motor.)

We performed simulations to determine whether non-processive
depolymerization is consistent with the experiments of Varga et al. In
this model (NP), we assumed that each motor could remove only one
tubulin dimer.  We tried but failed to match the experimentally
observed traces assuming completely stable MTs.  Increases in the
motor on rate by up to a factor of 1000 still produced simulations in
which depolymerization occurred much more slowly than seen in
experiments (data not shown).

Nonprocessive depolymerization by itself is insufficient to account
for the experimental results. However, combining model NP with
intrinsically unstable MTs is partially consistent with experiments.
The apparent motor processivity increases if protofilaments are not
completely stable.  We altered the protofilament depolymerization rate
$\beta$ to depend on the number of neighbors the terminal tubulin
dimer has. This modeling choice mimics the stabilizing lateral bonds
thought to be present between protofilaments
\cite{vanburen02,vanburen05}.  We note that our simple model does not
attempt to fully describe MT dynamics, and we considered only the
limiting case of strong lateral interactions: those protofilaments
missing at least one neighbor had a very high intrinsic
depolymerization rate $\beta_{0,1} = 13$ s$^{-1}$.  In this case,
model NP produced similar behavior to that seen in experiments only
with an on rate a full 20-fold higher than our estimate.  The maximum
depolymerization rate was then 2 dimers s$^{-1}$ or 1 $\mu$m min$^{-1}$.

From this analysis we conclude that if kinesin-8 motors depolymerize
MTs nonprocessively, GMPCPP-stabilized MTs must show some intrinsic
instability. Specifically, protofilament interactions must cause at
least one full row of dimers to be destabilized by the removal of a
single dimer.  Based on current experimental results, we cannot
unambiguously distinguish this case from a processive motor and stable
protofilaments. However, we note that the averaged length versus time
traces are not identical between these two cases.  There is less time
variation in the MT depolymerization rate in model NP than in model P.

The apparent motor processivity may be due to the inherent instability
of MT protofilaments lacking neighbors. However, our results indicate
that Kip3p likely depolymerizes MTs processively.

\subsection*{Fluctuations during processive depolymerization}

\begin{figure}[t]
  \centering
  \includegraphics[width=8cm]{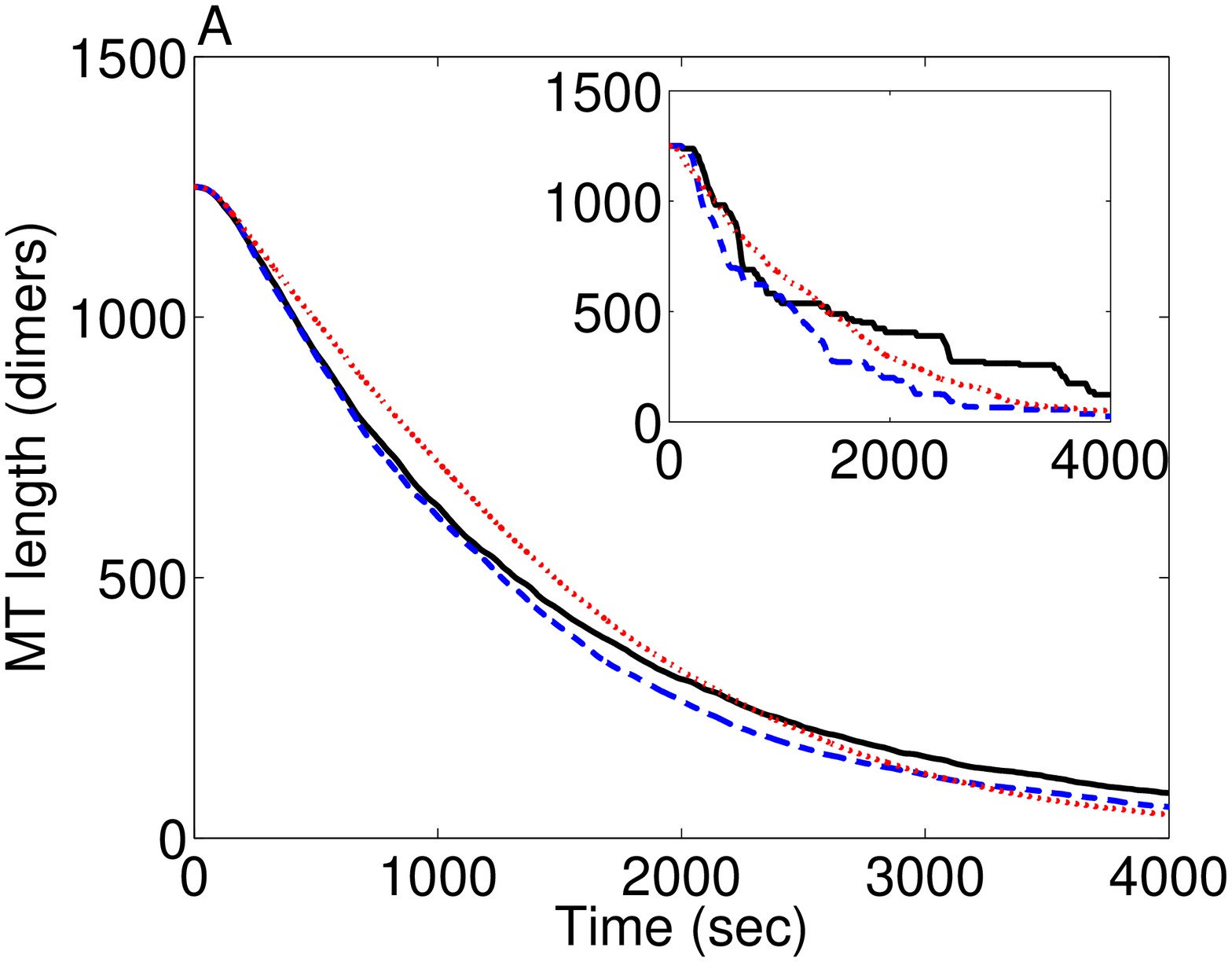}
  \includegraphics[width=8cm]{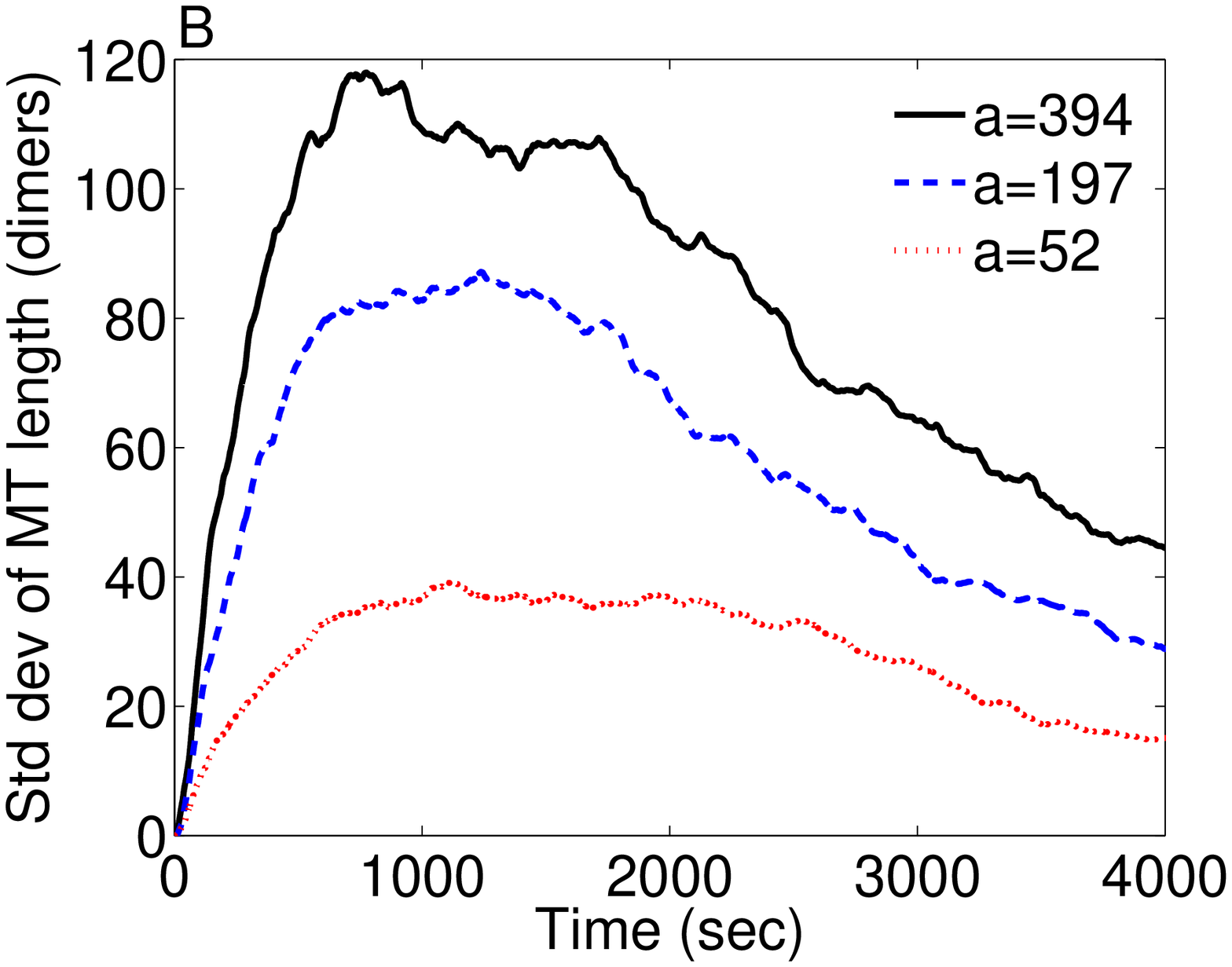}
  \includegraphics[width=8cm]{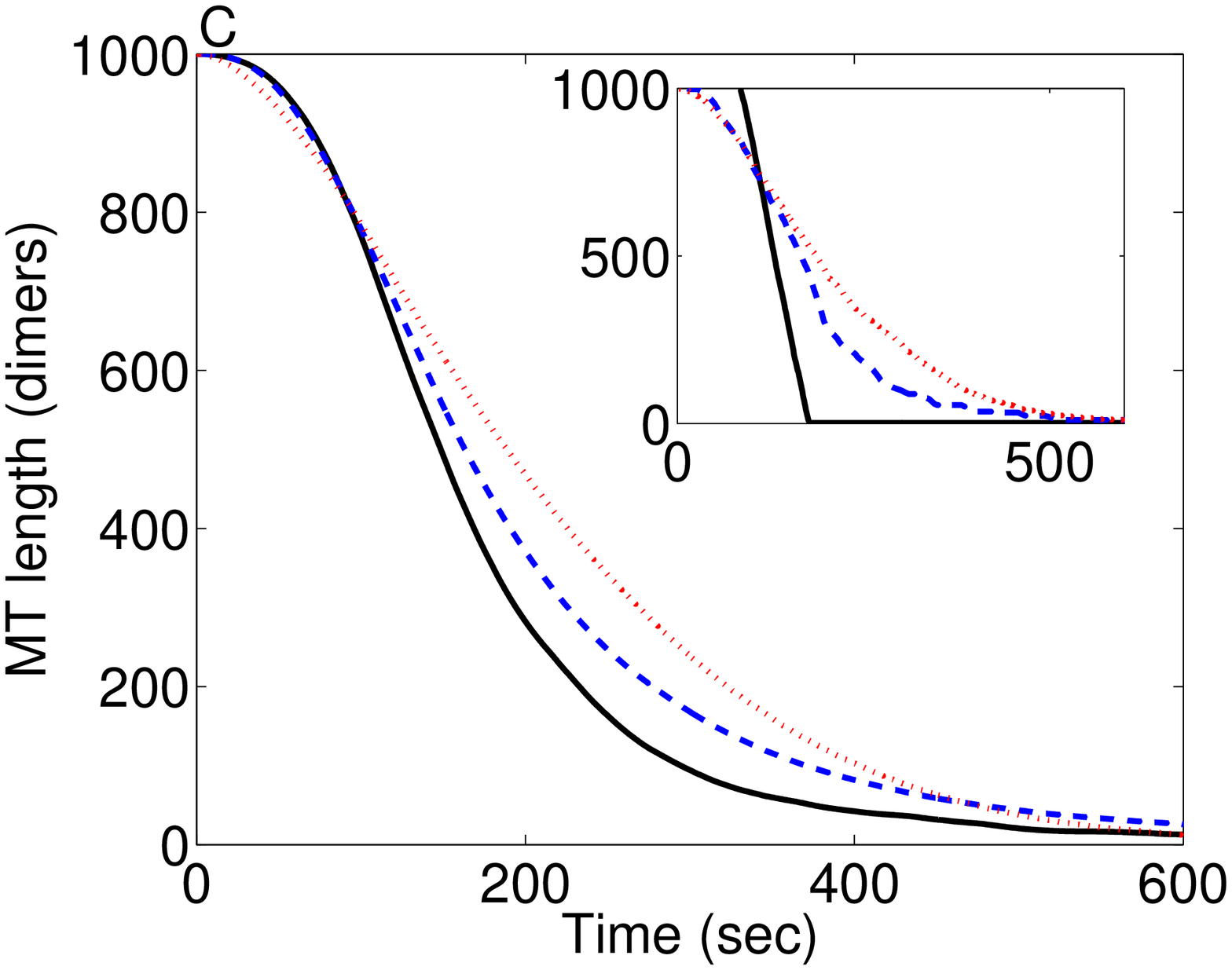}
  \includegraphics[width=8cm]{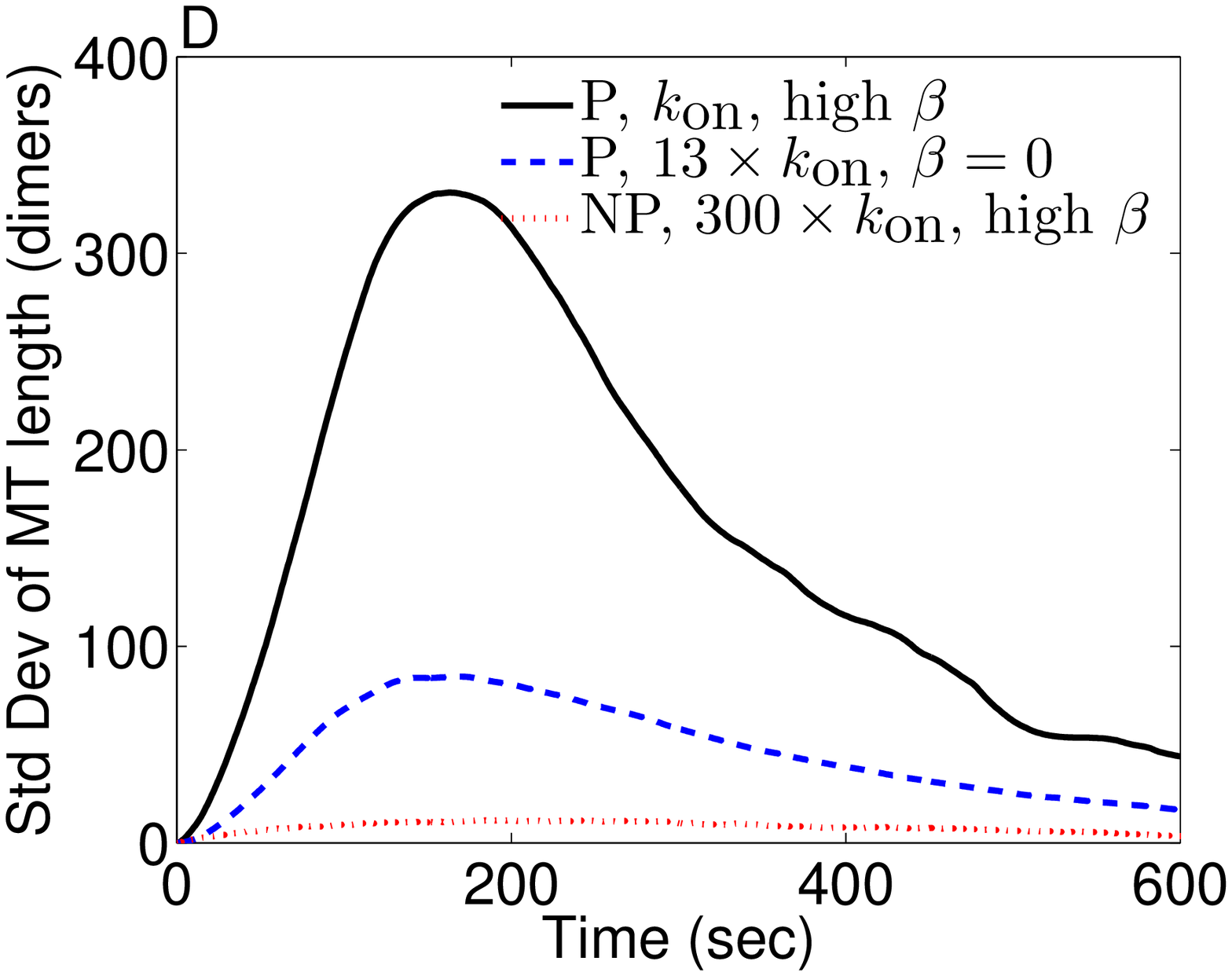}
  \caption{  }
  \label{variance}
\end{figure}

Our results so far have focused on averaged MT behavior. However,
fluctuations of MT length about the average are sensitive to kinesin-8
motor depolymerization processivity.  To illustrate this effect, we
chose parameters that give similar average length versus time dynamics
by keeping the product of the microscopic depolymerization rate
$k_-^{\mbox{\small end}}$ and the on rate constant $k_{\mbox{\small
    on}}$ approximately fixed while varying the maximum
depolymerization processivity, $a$ (fig.~\ref{variance}).  When $a$ is
increased, the dynamics of  MT length as a function of time become
more rough (fig.~\ref{variance}A, inset).  Motivated by the work of
Shaevitz \textit{et al.}, who considered the variance of stepping
behavior for conventional kinesin and showed that the variance
increased as the step size increased \cite{shaevitzBJ2005}, we
quantified the fluctuations in MT length as a function of time by
determining the standard deviation of MT length in our simulations
(figure \ref{variance}B).   For independent protofilaments, we
found that the standard deviation increases with $a$.  As expected,
the maximum standard deviation scales as $a^{1/2}$.  For these
simulations in particular, the maximum standard deviation of MT length
$\approx 6 a^{1/2}$.  

Thus, we propose that experimental measurement of the variance in MT
length as a function of time can be used to assess depolymerization
processivity. One experimental technique to measure the fluctuations
is to start with MTs of a certain average length, measure the MT
length again after a fixed time, and determine changes in the width of
the MT length distribution. Suppose the experimental parameters
correspond to those used in fig.~\ref{variance}A,B. Our idealized
simulation started with 10-$\mu$m long MTs; we then determined the
length distribution of MTs after approximately 1000 seconds of
depolymerization, when the variance is largest. The standard deviation
of MT lengths in the simulations was approximately 320 nm for $a=52$,
700 nm for $a=197$, and 950 nm for $a=394$, a difference that could be
measurable by high-resolution light microscopy.  A real experiment
would begin with a distribution of MT lengths, but would still observe
broadening of the MT length distribution. (Note that the precise value
of the standard deviation depends on the parameters used in the model.
Improved measurement of parameters such as the motor binding rate
would allow more accurate prediction of the fluctuations.)

In addition, the magnitude of the fluctuations depends on the strength
of lateral bonds between protofilaments in the MT. We considered
simulations without and with strong lateral interactions
($\beta_{0,1}=13$ s$^{-1}$), as shown in fig.~\ref{variance}C,D.

\section*{Mean-field model of motor-MT dynamics}

The Monte Carlo simulations accurately represent the full model, but
the equations are complex and analysis requires running many
simulations. We therefore developed simpler mean-field models that
describe the average occupancy of motors along the MT and the position
of the MT end.  A mean-field model considers averaged values of
variables such as motor occupancy and depolymerization rate, and
therefore neglects stochastic fluctuations.  While
these models are approximations to the full model, they are useful for
a simpler, approximate analysis and for understanding the importance
of fluctuations (which are absent in the mean-field models).

We analyzed two classes of mean-field model: ({\it i})  a description
of the motor occupancy profile, which characterizes the distribution of
motors along the MT, and ({\it ii}) a description of the MT end, which
uses the results of the motor occupancy studies to develop a simple
model of the MT depolymerization dynamics.

\subsection*{Motor occupancy profile}

To develop and analyze this model, we neglect MT dynamics.  (Recall
that in the \textit{in vitro} experiments, the MT intrinsic dynamics
primarily affect the long time behavior, i.e., when the tubulin
concentration in the flow chamber is sufficiently high to cause
significant polymerization.) The average fractional occupancy of
motors along the MT, $\rho$, is described by \cite{parmeggiani04}
\begin{equation}
  \frac{\partial \rho}{\partial t} = -v \frac{\partial \rho}{\partial x} +
  k_{\mbox{\small on}}c (1-\rho)- k_{\mbox{\small off}}\rho.  
\label{eq:rhox}
\end{equation}
On the right hand side, the first term represents the rate of change
of motor concentration due to biased motion of the motors with
velocity $v$, the second term represents the binding of motors to
unoccupied sites at rate $k_{\mbox{\small on}} c $, and the third term
represents unbinding of motors from occupied sites at rate
$k_{\mbox{\small off}}$.  The bulk motor concentration $c$ is assumed
constant. This equation treats motor crowding effects in a mean-field
approximation: the rate of binding of motors to the MT is assumed
proportional to $(1-\rho)$, decreasing in proportion to the average
occupancy of a particular site. We neglect crowding effects in the
transport term $\sim v \frac{\partial \rho}{\partial x}$, which if
considered make the density equation nonlinear~\cite{parmeggiani04}.

The steady-state density distribution away from either of the MT ends
is given by the constant solution to this equation: 
\begin{equation}
  \rho = \rho_0 = \frac{ k_{\mbox{\small on}}c }{  k_{\mbox{\small off}}
    + k_{\mbox{\small on}}c }.
  \label{eq:rhoo}
\end{equation}
Note that if the on rate is sufficiently small, that is,
$k_{\mbox{\small on}}c \ll k_{\mbox{\small off}}$, then $\rho_0
\approx k_{\mbox{\small on}}c/k_{\mbox{\small off}} = c/K$. In other
words, the average motor occupancy on the MT is approximately the bulk
motor concentration divided by the equilibrium constant for motor
binding to the MT.  

\begin{figure}[t] 
      \centering
      \includegraphics[width=10 cm]{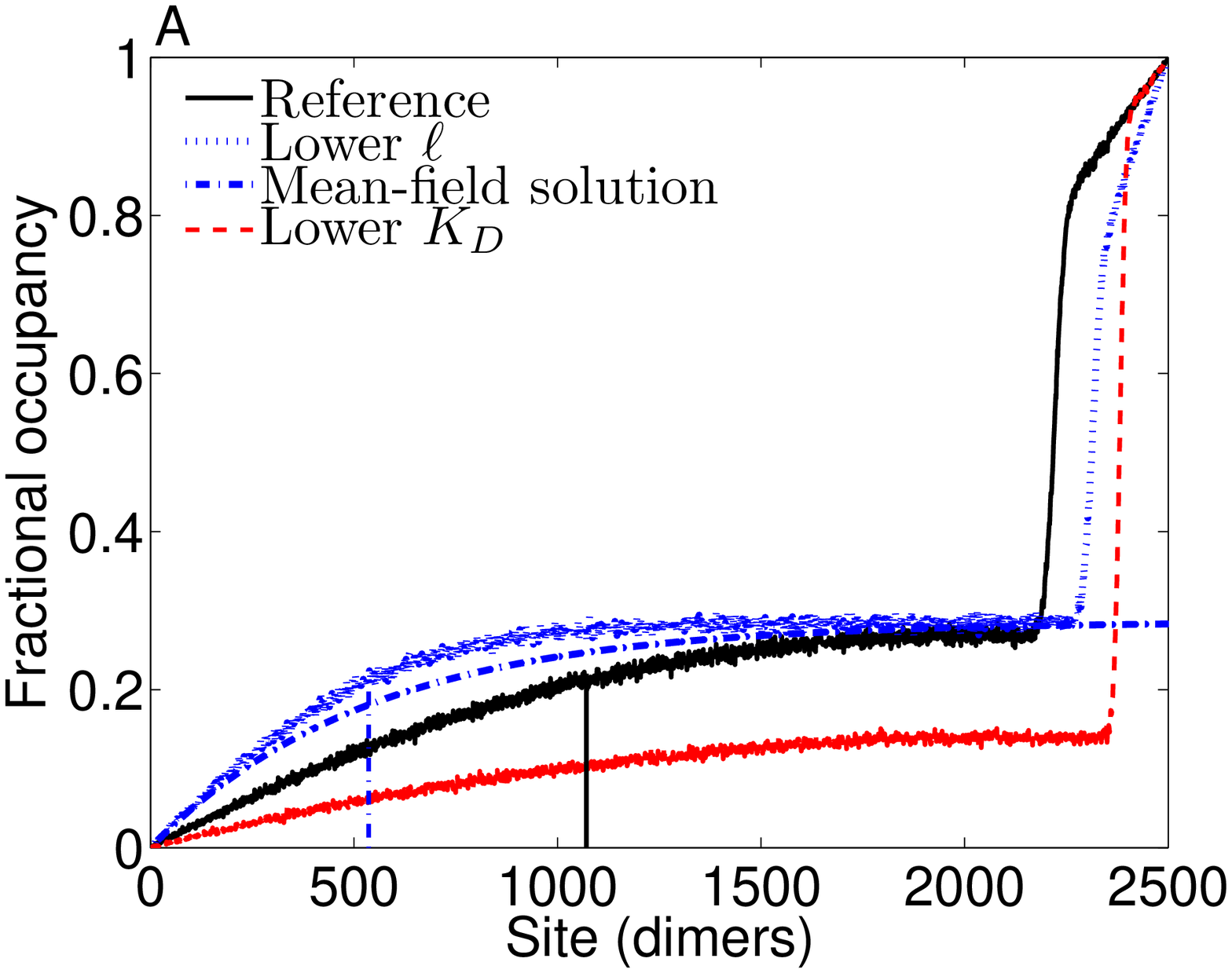}
      \includegraphics[width=10 cm]{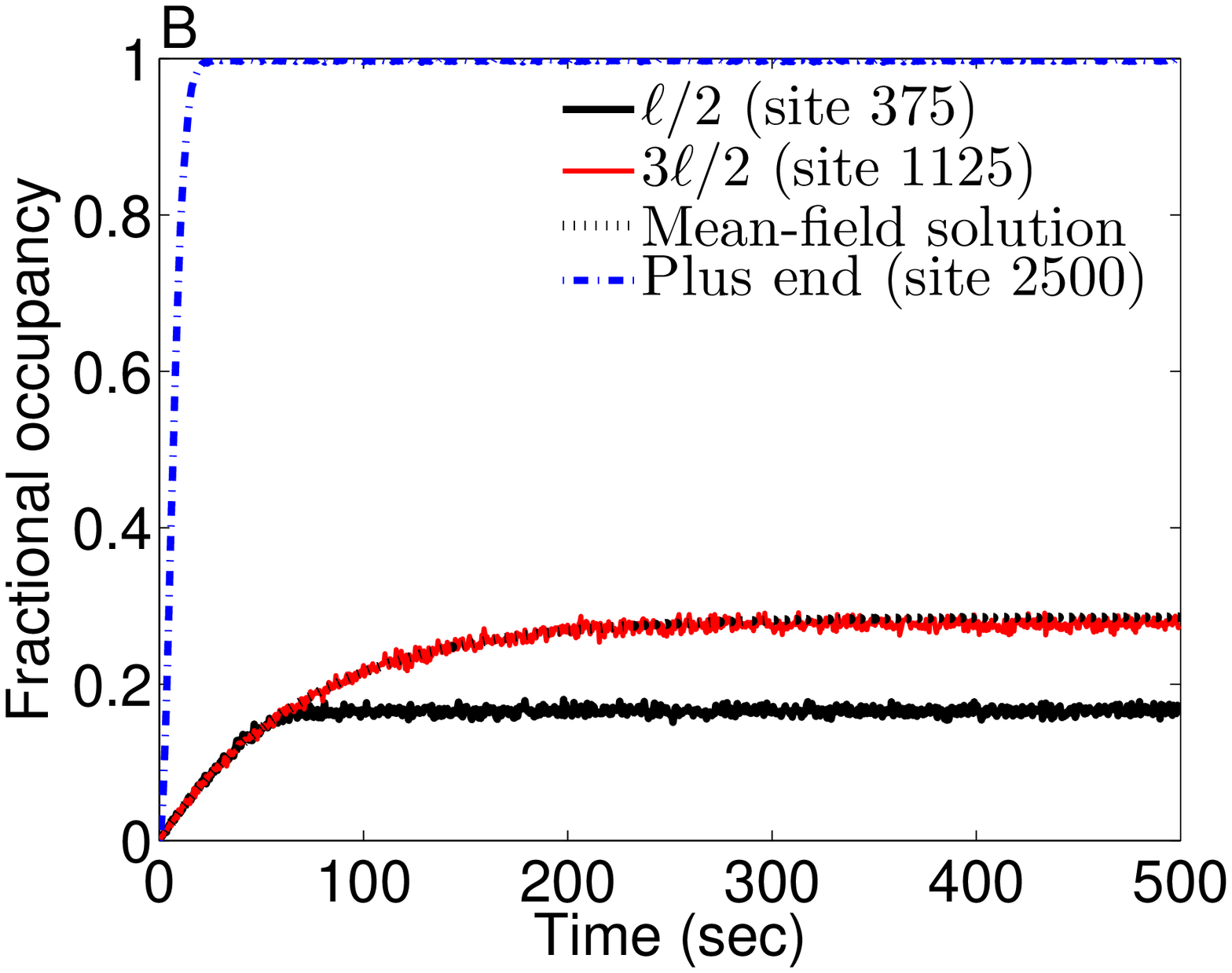}
      \caption {  }
      \label{SSdensity}
    \end{figure}

Away from either of the MT ends, the density approaches the constant
value $\rho_0$. If we consider a spatially constant occupancy which is
not equal to $\rho_0$, the time dependence of eqn.~\eqref{eq:rhox} has
exponential solutions. If at time $t=0$ motors are introduced to the
system, the density far from the MT ends will change in time according
to\begin{equation}\rho(t) = \rho_0 (1-e^{-t/\tau}).\label{eq:rhotau}
\end{equation}
The characteristic time scale is
\begin{equation}
\tau = \frac{1}{
k_{\mbox{\small off}} + k_{\mbox{\small on}}c}.
\label{eq:tau}
\end{equation}
As shown in figure \ref{SSdensity}, this expression agrees well with
the simulation results, giving a value of the occupancy within 10\% of
the value from simulations for the region away from the plus end of
the MT.

Near the MT ends there is a ``boundary layer'' where transport effects
and boundary conditions change the motor density away from $\rho_0$
\cite{parmeggiani04,nowak07,govindan08}. Near the minus end (small
$x$) the boundary layer has low motor density; this occurs because the
number of motors that have moved out of a region near the minus end is
not balanced by motors moving into that region from the minus end. At
$x=0$ the motor density is exactly zero.  In the linear mean-field
approximation described by equation \eqref{eq:rhoo}, the steady-state
density is then
\begin{equation}
\rho(x) = \rho_0 ( 1- e^{-x/\lambda}),
  \label{eq:rho}
\end{equation}
where $\lambda = v/( k_{\mbox{\small off}} + k_{\mbox{\small on}}c) =
v \tau$ is the length scale that characterizes the boundary layer of
the density distribution near the minus end. In figure
\ref{SSdensity}, we show that the steady-state simulation results are
well represented by this occupancy profile.

For small on rate ($k_{\mbox{\small on}}c \ll k_{\mbox{\small off}}$)
this length scale is $\lambda \approx v/k_{\mbox{\small off}} = \ell$,
the motor run length. For $x \ll \lambda$, the motor occupancy
increases linearly with $x$, with slope $\rho_0/\lambda =
k_{\mbox{\small on}}c/v$.

Motor occupancy profiles consistent with this result have been observed
both \textit{in vitro} and \textit{in vivo}. Varga et al.~observed
a linear increase in Kip3p motor occupancy with position \textit{in
  vitro} \cite{VargaNCB2006}.  (We note that if the slope of this
linear increase could be measured it would provide a direct measure of
the motor on rate $k_{\mbox{\small on}}$.)  Stumpff et al.~imaged
human kinesin-8 fluorescence along MTs in fixed cells and observed a
gradient in fluorescence which is qualitatively consistent with the
model predictions \cite{stumpff08}; similar gradients in motor
fluorescence were seen in yeast cells by Varga et al.~and Gupta et al.
\cite{VargaNCB2006,GuptaNCB2006}.

\subsection*{Mean-field model of plus-end dynamics}

The dynamics of the microtubule are controlled by the density of
motors at the MT end.  Here we formulate and analyze a mean-field
description of the motor occupancy at the MT end and the MT length.
Because motors promote MT shortening, the density at the MT plus end
is constantly changing, making the dynamics here typically non-steady
state.

We define $\rho_e(t)$ to be the average motor occupancy at the last
site on the MT end. In this model we do not consider protofilament
interactions, so we are effectively considering a single-protofilament
MT.  The MT length is represented by $L(t)$ and its rate of change is
$dL/dt$, which is negative during depolymerization. The coupled
dynamics of the end occupancy and MT end change according to:
\begin{eqnarray}
  \frac{d \rho_e}{d t} &=& \left(v-\frac{d L}{d
      t}\right)\rho(L-\epsilon,t)(1-\rho_e)   
  - k_{\mbox{\small off}}^{\mbox{\small end}} \rho_e,  \label{eq:dndt} \\
  \frac{d L}{d t} &=&  - k_-^{\mbox{\small end}} \rho_e.
\label{eq:dldt}
\end{eqnarray}
The first term on the right side of eqn.~\eqref{eq:dndt} represents
the arrival of motors from the region adjacent to the end, where the
density is $\rho(L-\epsilon,t)$ and $\epsilon$ is a small parameter.
Solving for the full time dependence of $\rho(L-\epsilon,t)$ is, in
general, difficult; one would have to solve for the time varying
density near a moving boundary.  However, we have assumed that a motor
falls off while depolymerizing if the dimer directly adjacent is
occupied (fig.~\ref{model}C).  Thus, if a transient clump of motors
developed at the MT plus end, it would be quickly removed at rate
$k_-^{\mbox{\small end}}=13$ s$^{-1}$, faster than other
processes in the model.  Thus, we can approximate
$\rho(L-\epsilon,t)\approx \rho(L-\epsilon)\approx\rho(L)$, where
$\rho(x)$ is the motor occupancy for a region far from the MT plus
end.  This density $\rho(x)$ may vary in time or be a steady-state
value, but we assume that $\rho(x)$ is controlled by dynamics away
from the MT end.  The second term on the right side of
eqn.~\eqref{eq:dndt} describes unbinding of the motor at the end.  In
eqn.~\eqref{eq:dldt}, we assume that the rate of MT shortening is
proportional to the depolymerization rate and the motor density at the
end. Note that if $\rho_e$ is constant in time, then according to
eqn.~\eqref{eq:dldt} the MT shortens at a constant rate.

These equations can be combined to write
\begin{equation}
  \frac{d \rho_e}{d t} = \left(v+k_-^{\mbox{\small end}}
    \rho_e\right)\rho(L)(1-\rho_e)  - k_{\mbox{\small
      off}}^{\mbox{\small end}} \rho_e. 
\label{eq:rhoe}
\end{equation}

Equation \eqref{eq:rhoe} can be numerically integrated or studied
analytically within certain limits. Below we first determine
analytically the constant depolymerization rate of very long MTs,
where $\rho(L) \approx \rho_0$ as determined from
eqn.~\eqref{eq:rhoo}. We then compare this predicted constant
depolymerization rate with simulations on long MTs.

Then we will consider how long it takes to approach this constant
depolymerization rate, and find a typical time scale of tens of
seconds. This rapid approach to constant depolymerization led us to
consider quasi-static depolymerization. We will assume that even when
the motor density away from the MT end varies spatially, the motor
occupancy at the MT plus end rapidly tracks these changes.

\subsubsection*{Constant depolymerization of long MTs}

\begin{figure}
    \centering 
   \includegraphics[width=12 cm]{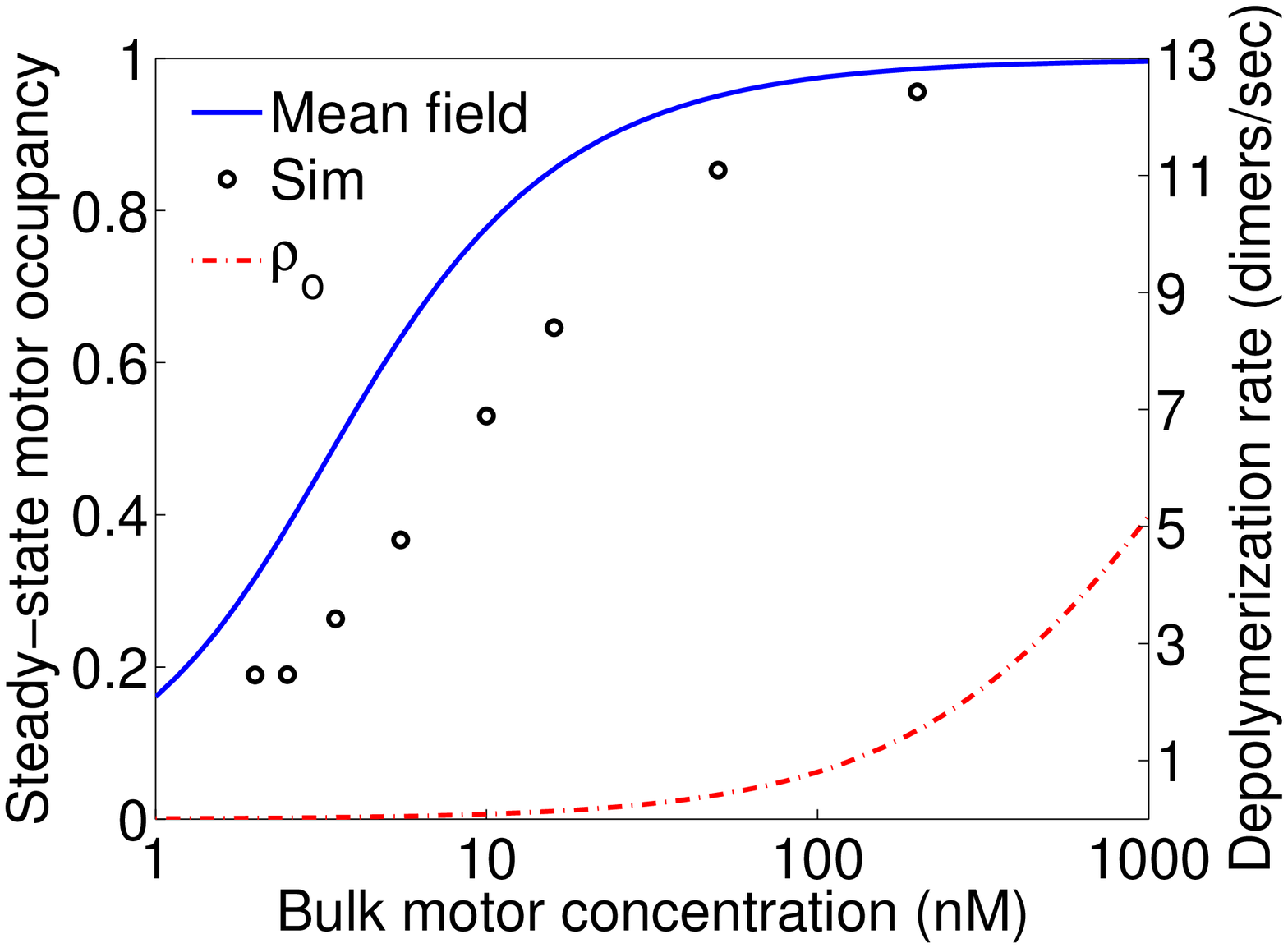}
   \caption {}
  \label{shortenss} 
\end{figure}

We will consider first the limit of long MTs with a constant motor
density, so that $\rho(L) = \rho_0$, independent of MT length. In this
limit, the constant depolymerization velocity of the MT is determined
by the steady-state value of $\rho_e$.  In this case, equation
\eqref{eq:rhoe} simplifies to one with no $L$ dependence:
\begin{eqnarray}
  \frac{d \rho_e}{d t} &=& \left(v+k_-^{\mbox{\small end}}
    \rho_e\right)\rho_0(1-\rho_e)  - k_{\mbox{\small
      off}}^{\mbox{\small end}} \rho_e,\\
 &=&-k_-^{\mbox{\small end}}\rho_0 \rho_e^2 +
  [(k_-^{\mbox{\small end}}-v) \rho_0 -  
  k_{\mbox{\small off}}^{\mbox{\small end}}] \rho_e + v \rho_0.
\label{eq:rhoelong}
\end{eqnarray}
This equation has steady state solutions determined by the quadratic
equation
\begin{equation}
  k_-^{\mbox{\small end}}\rho_0 \rho_e^2 + 
  [(v-k_-^{\mbox{\small end}}) \rho_0 +  
  k_{\mbox{\small off}}^{\mbox{\small end}}] \rho_e - v \rho_0 = 0. 
\label{eq:rhoessA}
\end{equation}
Defining $g=(v-k_-^{\mbox{\small end}}) \rho_0 + k_{\mbox{\small
    off}}^{\mbox{\small end}}$ we can write the solutions as
\begin{equation}
 \rho_{e\pm} = \frac{-g \pm \sqrt{g^2+4 v k_-^{\mbox{\small end}}
     \rho_0^2}}{2 k_-^{\mbox{\small end}} \rho_0}.
\label{eq:rhoess}
\end{equation}
The physically relevant solution (with $0 \le \rho_e \le 1$) is the
negative root. When the end occupancy is constant, the
depolymerization rate is $|dL/dt| = k_-^{\mbox{\small end}} \rho_e$. In
fig.~\ref{shortenss} we show the predicted steady-state occupancy at
the MT end, the resulting shortening rate, and comparison with
simulations. (For details of the simulations, see the section on
length-dependent depolymerization below.)

The effect of fluctuations is to decrease the depolymerization rate,
relative to mean-field predictions. This is intuitively reasonable,
because when a fluctuation leads to a higher than average density, the
motor at the MT end is then rapidly knocked off (see
fig.~\ref{model}), which leads to a decrease in the density at the
end. Therefore fluctuations which decrease the motor density at the
end decrease the depolymerization rate more than fluctuations which
increase the motor density at the end increase the depolymerization
rate. Despite this small error, the mean-field theory predicts the shape of
the curve correctly and determines the depolymerization rate to within
50\%.

\subsubsection*{Approach to steady state}

\begin{figure}
    \centering 
   \includegraphics[width=10 cm]{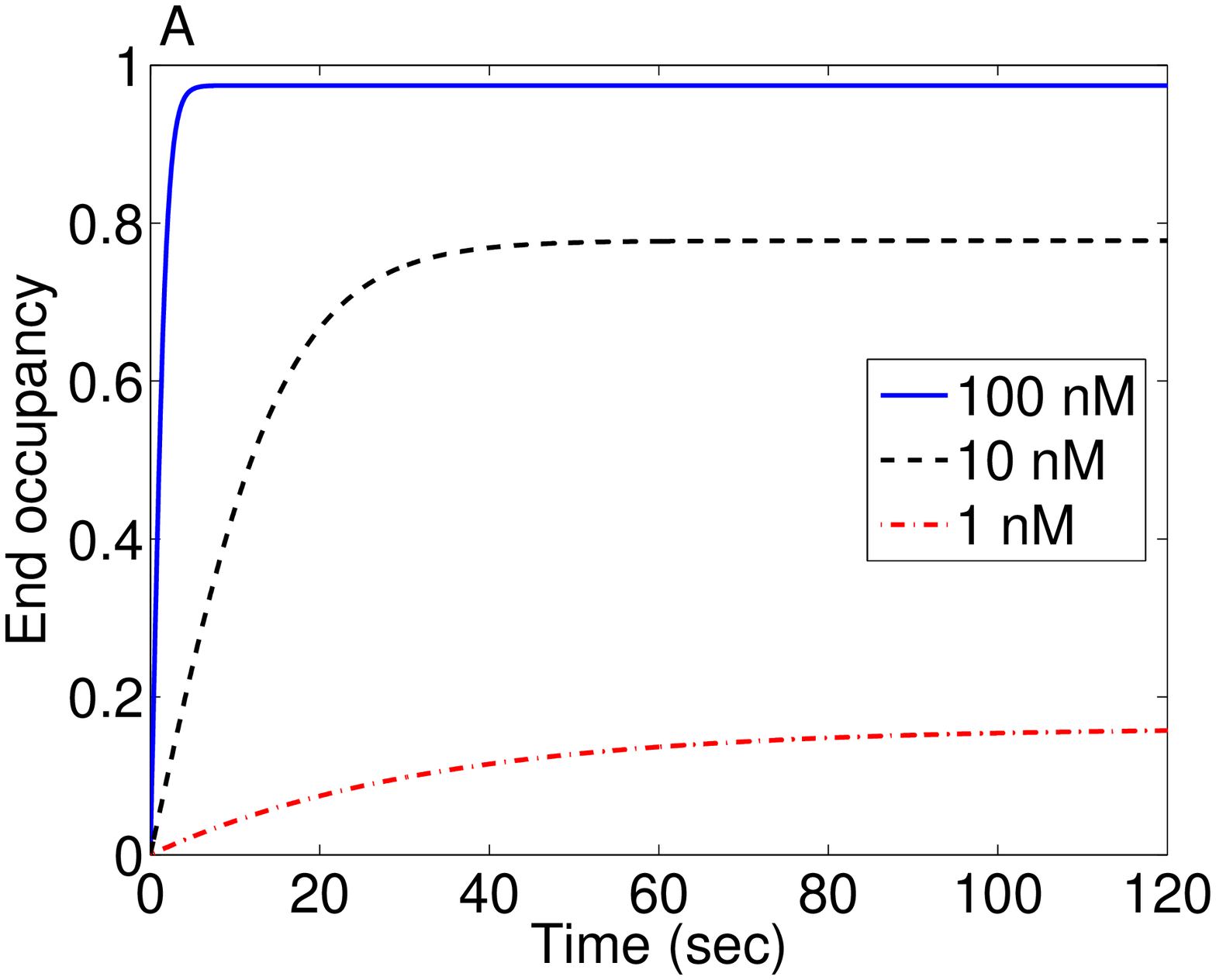}
   \includegraphics[width=10 cm]{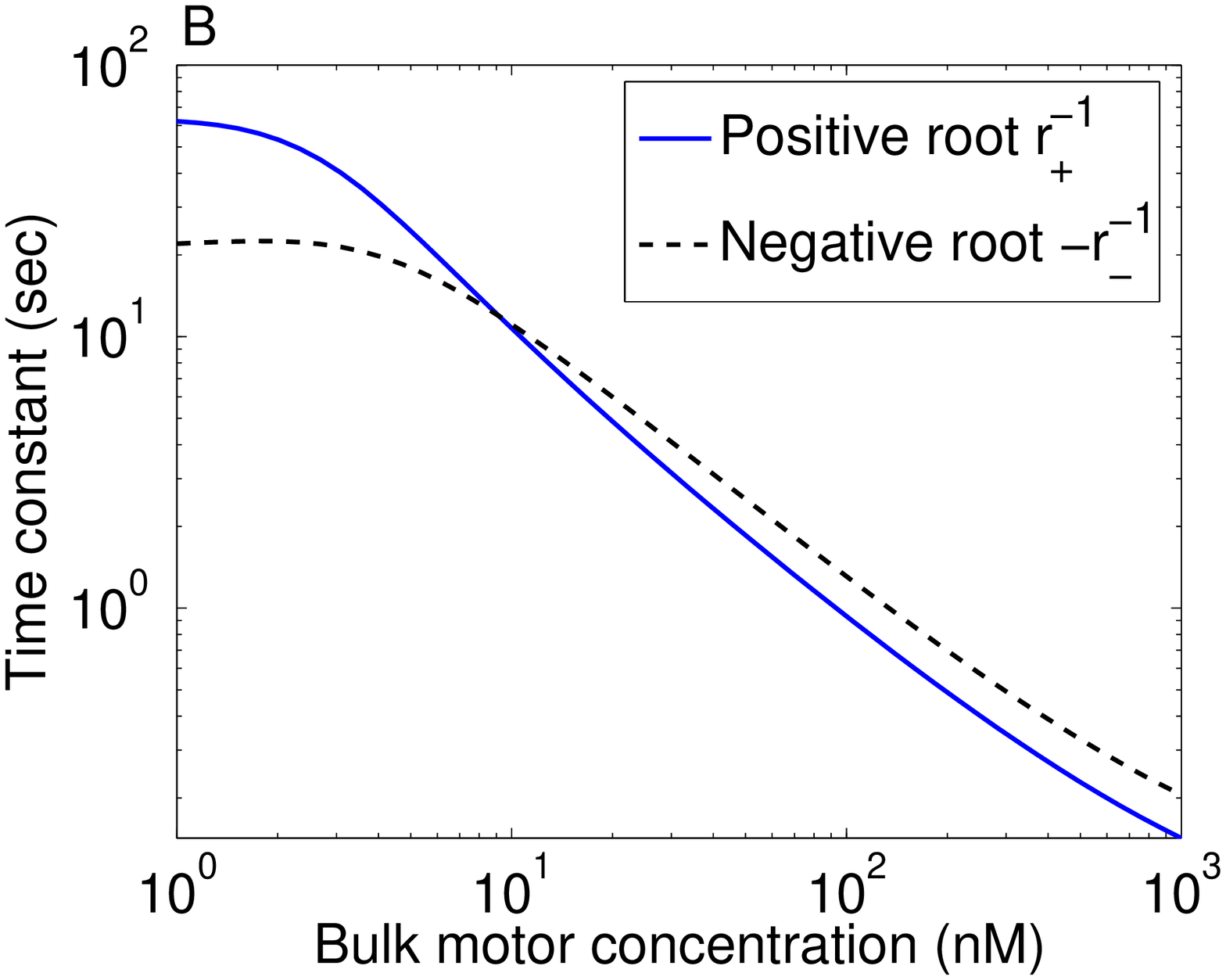}
   \caption {}
  \label{timeconst} 
\end{figure}

We note that these predictions assume that the MTs start with an
initial length long enough that the steady-state motor occupancy at
the end can be reached. This assumption may not apply in experiments
(see below). We therefore determine the approach of solutions of
eqn.~\eqref{eq:rhoelong} to steady state. This equation is a Ricatti
equation with constant coefficients, of the form $\dot{\rho} = f
\rho^2 + g \rho + h$, which can be transformed into a linear, 2nd
order ODE using the substitution $u(t) = \exp[ -\int f \rho(t) dt]$.
The function $u(t)$ then has two exponential solutions $u(t) \sim
e^{r_{\pm} t}$ with inverse time constants
\begin{equation}
r_{\pm} = \frac{1}{2} \left( g \pm \sqrt{g^2-4fh} \right).   
\label{eq:rpm}
\end{equation}
The resulting time-dependent solution for the density, given
$\rho_e(t=0)=\rho_i$, is
\begin{equation}
  \rho_e(t) = \frac{1}{k_-^{\mbox{\small end}} \rho_0} \left( 
    \frac{r_+(r_-+ k_-^{\mbox{\small end}} \rho_0 \rho_i) e^{r_+t} - r_-
      (r_++ k_-^{\mbox{\small end}} \rho_0 \rho_i) e^{r_-t}}
    {(r_-+ k_-^{\mbox{\small end}} \rho_0 \rho_i) e^{r_+t} - 
      (r_++ k_-^{\mbox{\small end}} \rho_0 \rho_i) e^{r_-t}}  
  \right).   
\label{eq:rhotA}
\end{equation}
Typical values of parameters give $h<0$ and therefore $r_+>0$ and
$r_-<0$, so the dynamics will be controlled by the $r_+$ terms for
long times.  The decay times $r_{\pm}^{-1}$ are of order seconds to
tens of seconds, with slower times for lower concentrations. See
fig.~\ref{timeconst} for the values of the time constants and a
typical time trace determined by eqn.~\eqref{eq:rhotA}.

Note that this equation assumes that the motor density outside the end
of the MT has reached the steady-state value $\rho_0$. As shown in
fig.~\ref{SSdensity}, it can take several minutes for the density
(away from the MT end) to equilibrate. Therefore the dynamics of the
MT end are limited by the dynamics of motors away from the end more
than by the processes at the end.

\subsubsection*{Quasi-static depolymerization of MTs}

Outside the regime of constant depolymerization, we can still make a
simple approximation to the depolymerization rate by assuming that the
plus end of the MT has a motor density determined by the instantaneous
solution of the steady-state equation, but with a varying density away
from the end. In other words, we assume that the time for motor
density at the MT plus end to reach steady state is short compared to
other time scales in the problem. This is a reasonable approximation,
given that the dynamics at the end of the MT reach steady state in
seconds to tens of seconds (fig.~\ref{timeconst}) while MT shortening
typically takes minutes.

In the quasi-static approximation, we solve
\begin{eqnarray}
  0 &=& \left(v-\frac{d L}{d t}\right)\rho(L)(1-\rho_e)  
  - k_{\mbox{\small off}}^{\mbox{\small end}} \rho_e,  \label{eq:dndtqs} \\
  \frac{d L}{d t} &=&  - k_-^{\mbox{\small end}} \rho_e.
\label{eq:dldtqs}
\end{eqnarray}
The solutions here are similar to those of eqn.~\eqref{eq:rhoess}, but
with the varying density $\rho(L)$ apparent in the solution:
\begin{equation}
  k_-^{\mbox{\small end}}\rho(L) \rho_e^2 + 
  [(v-k_-^{\mbox{\small end}}) \rho(L) +  
  k_{\mbox{\small off}}^{\mbox{\small end}}] \rho_e - v \rho(L) = 0. 
\label{eq:qs}
\end{equation}
Defining $g(L)=(v-k_-^{\mbox{\small end}}) \rho(L) + k_{\mbox{\small
    off}}^{\mbox{\small end}}$, the physically relevant solution is
\begin{equation}
 \rho_{qs}(L) = \frac{-g(L) - \sqrt{g(L)^2+4 v k_-^{\mbox{\small end}}
     \rho(L)^2}}{2 k_-^{\mbox{\small end}} \rho(L)}.
\label{eq:rhoeqs}
\end{equation}
This quasi-static density at the end of the MT then determines the
depolymerization rate via eqn.~\eqref{eq:dldtqs}. We will study the
accuracy of this quasi-static approximation below in the section on
the phase diagram of length-dependent depolymerization.

\section*{Length-dependent depolymerization}

A key feature of the Varga et al.~results is the observation of
length-dependent MT depolymerization \cite{VargaNCB2006}, where the
depolymerization rate decreases as the MT length decreases. Since the
depolymerization rate cannot increase indefinitely with MT length, a
long MT ($L \gg \ell$) will not show length-dependent depolymerization
until it becomes sufficiently short.  The ``crossover'' length $d$ is
the length at which length-dependent depolymerization begins; for MT
lengths $L > d$ the depolymerization rate is constant while for $L < d$
the depolymerization rate decreases as $L$ decreases. 
Here
we determine the crossover length $d$ as a function of motor
parameters and experimental conditions.

Motors accumulate at the MT plus end for several reasons.  First,
motors have different residence times at the MT end than away from the
end.  Away from the ends of the MT, a motor moves away from a dimer at
rate $v$. The off rate for a motor from the last dimer along the MT is
$k_{\mbox{\small off}}^{\mbox{\small end}}$.  This will increase the
motor occupancy by a factor of approximately $ v/k_{\mbox{\small
    off}}^{\mbox{\small end}} \sim 200$ (although note that the
average motor occupancy per site cannot be $>1$; see
fig.~\ref{shortenss}).  Second, depolymerization moves the MT end
closer to motors away from the end; in the frame of the MT end motors
approach at a rate $v - dL/dt$, where $dL/dt$ is the rate of change of
MT length.  Third, motors could accumulate at the MT end due to direct
binding of motors to the end.  The latter effect is neglected here.

\begin{figure}[t] 
    \centering 
  \includegraphics[width=10 cm]{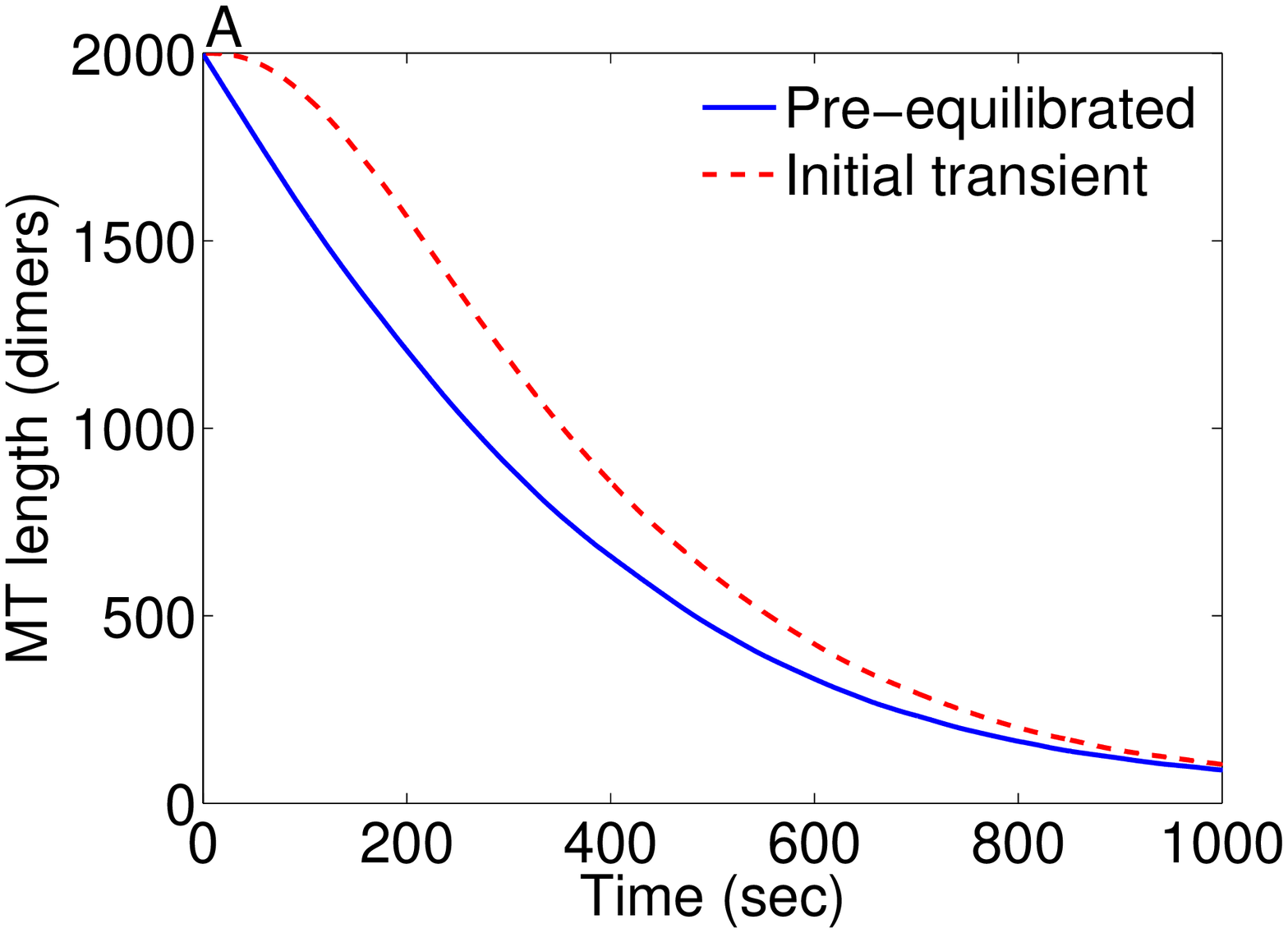}
  \includegraphics[width=10 cm]{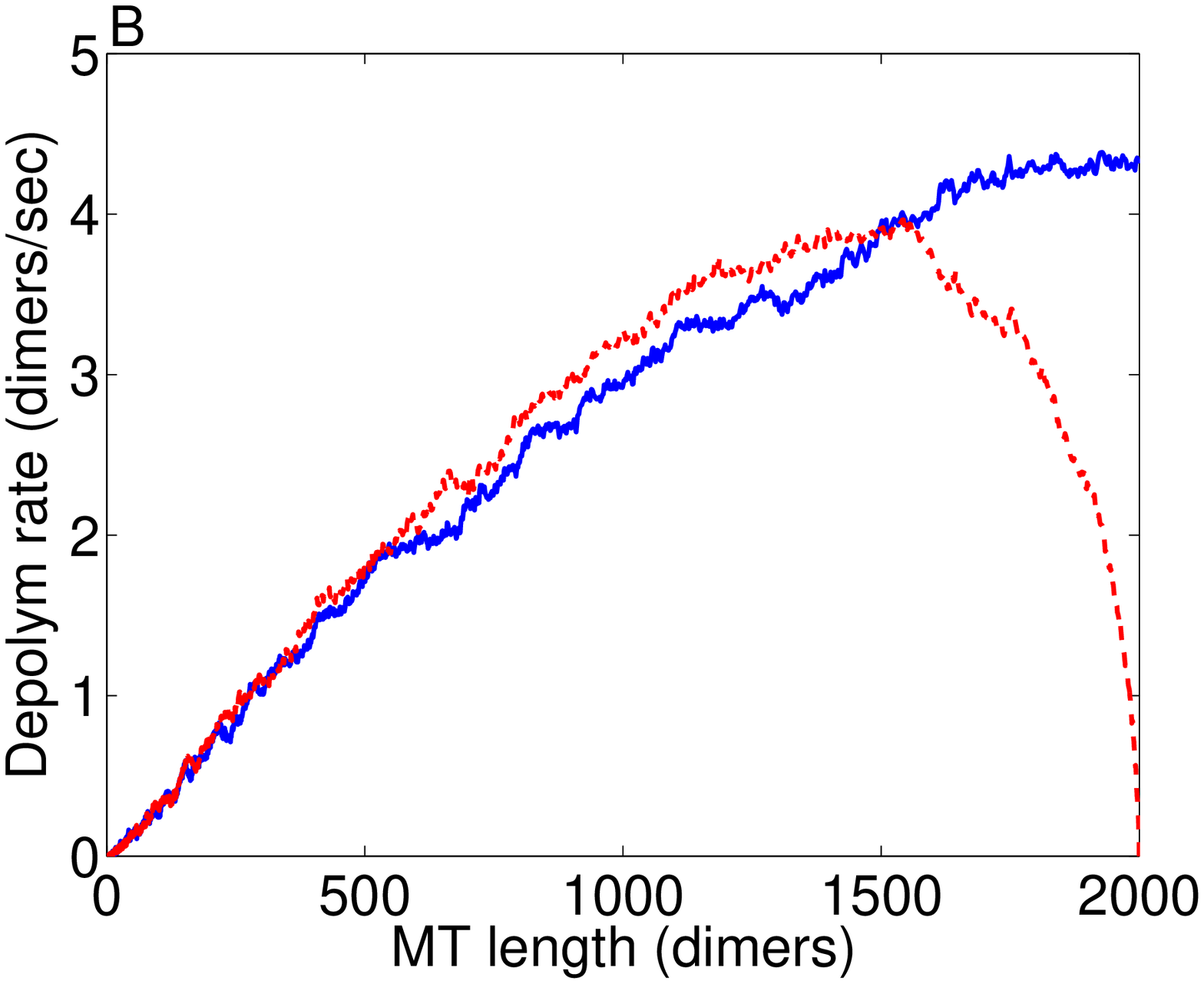}
  \caption {}
\label{equil-trans}
 \end{figure}

To characterize length-dependent depolymerization and to understand
its role in vivo, we estimated the crossover length $d$ analytically
and compare it to simulations.  We consider length-dependent
depolymerization in two limits: first we consider the case most
independent of initial conditions, which corresponds to starting with
very long MTs or pre-equilibrated motor occupancy along the MT.
Second, we consider the case where the initial motor occupancy on the
MT is zero; this case is more relevant to an experiment started with
the addition of motors to previously unoccupied MTs. See
fig.~\ref{equil-trans} for a comparison of depolymerization of
pre-equilibrated MTs to depolymerization with an initial transient.

\subsection*{Length-dependent depolymerization independent of initial conditions}

Here we characterize length-dependent depolymerization in a regime
which is, as much as possible, independent of the initial conditions
or the starting time of an experiment. Such a limit would be reached
in experiments (or in cellular conditions) if one either ({\it i})
started the experiment with very long MTs, or ({\it ii}) equilibrated
the motor density on the MTs before the start of depolymerization.

\begin{figure}
    \centering 
  \includegraphics[width=8 cm]{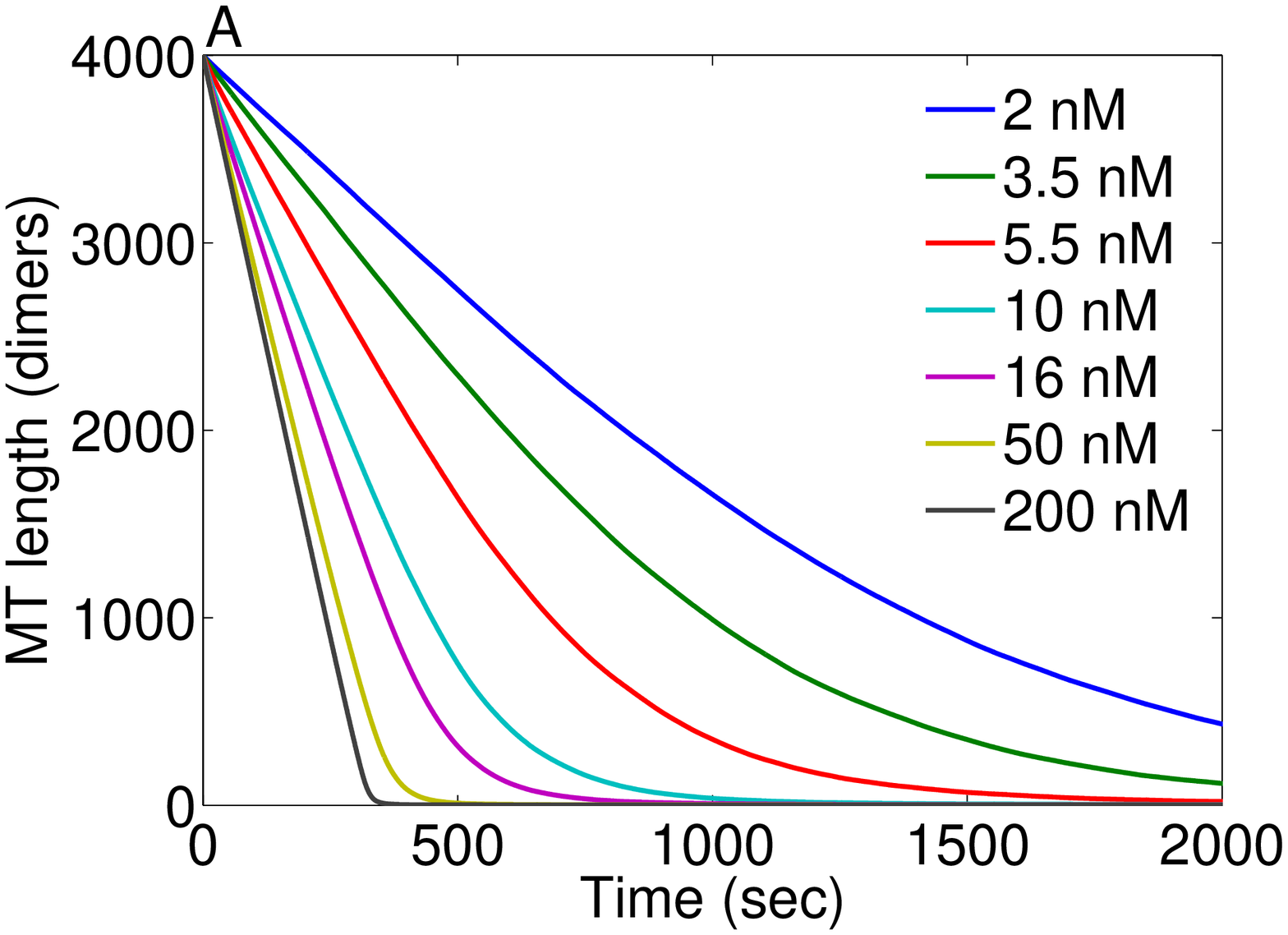}
  \includegraphics[width=8 cm]{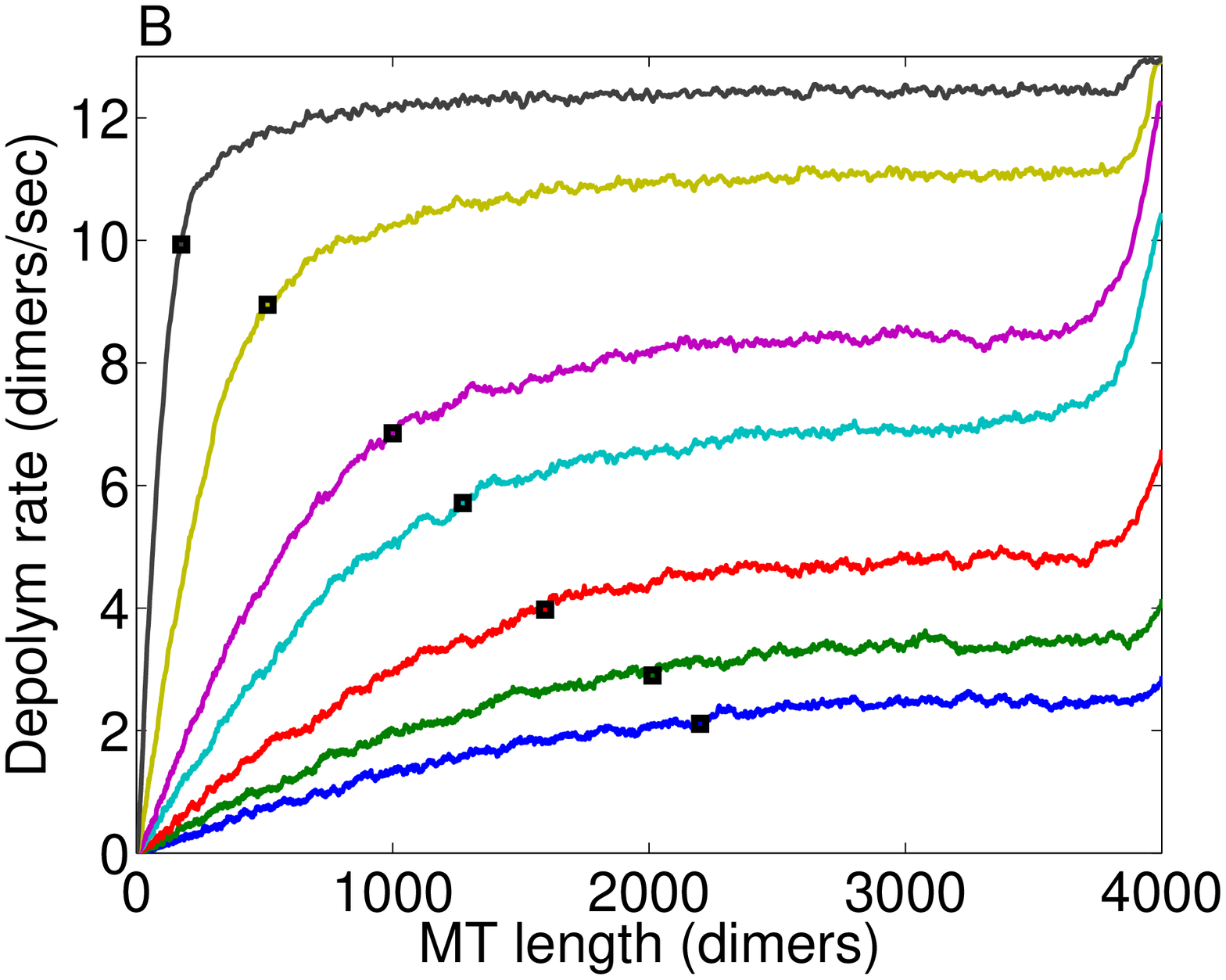}
  \includegraphics[width=8 cm]{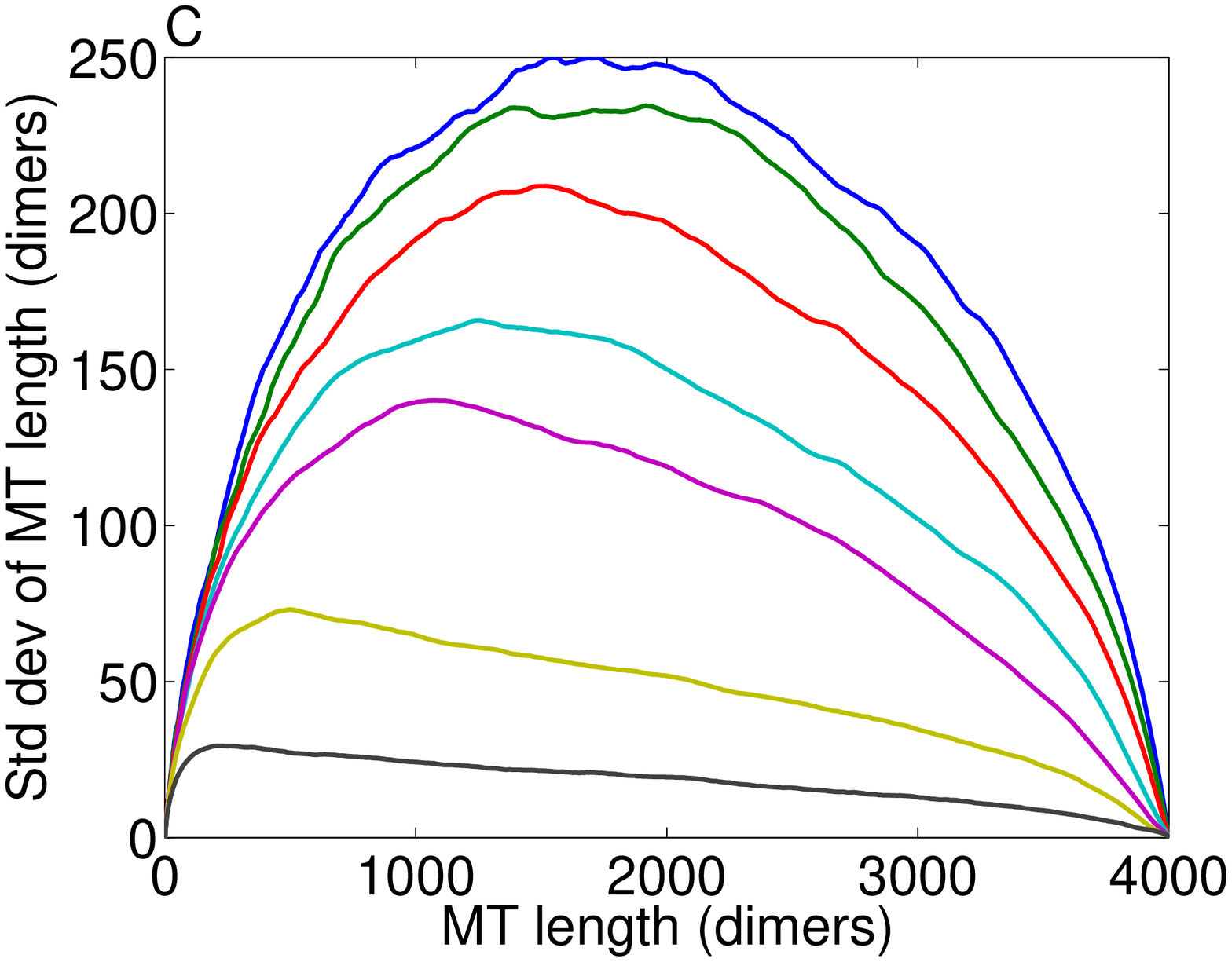}
  \includegraphics[width=8 cm]{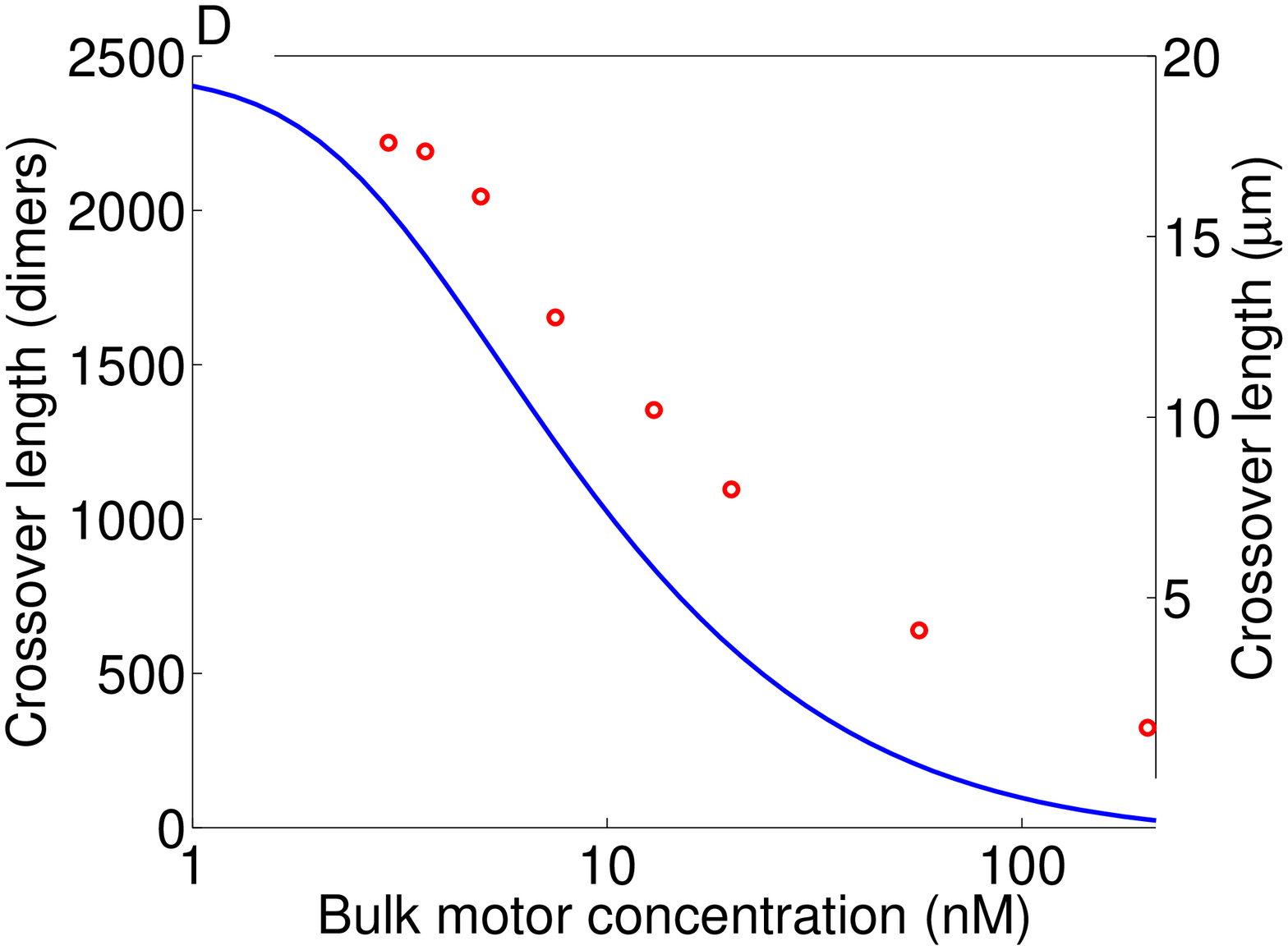}
  \caption {}
  \label{shorten} 
\end{figure}

To approach this limit in the simulations, we implemented both long
initial MTs and motor pre-equilibration. We started with MTs 4000
dimers long (32 $\mu$m) and ran the simulations for 1000 seconds with
motor binding, unbinding, and motion allowed but depolymerization
turned off. This starting point was then used for simulations of MT
shortening.  Averaged simulation results are shown in
fig.~\ref{shorten}. By examining the depolymerization rate versus MT
length (fig.~\ref{shorten}B), we can see that initially the
depolymerization rate is high, but it drops quickly as the clump of
motors at the MT plus end is removed during depolymerization. Then the
depolymerization rate reaches a constant value that holds over a range
of MT lengths. (This constant rate is compared to the predictions of
the mean-field model in fig.~\ref{shortenss}.) Once the MT is
sufficiently short, the depolymerization rate drops below the
steady-state value. Eventually the depolymerization rate drops to zero
for a MT of length zero.

We also determined the fluctuations in MT length about the average:
fig.~\ref{shorten}C shows the standard deviation in MT length as a
function of MT length during shortening. Lower motor concentration
leads to a significantly larger standard deviation in MT length,
indicating that larger fluctuations about the average occur when fewer
motors are present.

The length where the depolymerization rate drops below the
steady-state value is the ``crossover length'' $d$ where
length-dependent depolymerization sets in (fig.~\ref{shorten}D).  We
defined the crossover length $d$ as the MT length where the
depolymerization rate first decreased by 20\% from the steady-state
value. The results from simulations are shown in fig.~\ref{shorten}D.
We calculated $d$ in the mean-field model by rewriting
eq.~\eqref{eq:rhoe} using $L$ rather than $t$ as the independent
variable:
\begin{eqnarray}
  \frac{d \rho_e}{d L} \frac{dL}{dt} &=& \left(v+k_-^{\mbox{\small end}}
    \rho_e\right)\rho(L)(1-\rho_e)  - k_{\mbox{\small
      off}}^{\mbox{\small end}} \rho_e,\\
  \frac{d \rho_e}{d L} &=& \frac{-1}{k_-^{\mbox{\small end}} \rho_e } \left[
    \left(v+k_-^{\mbox{\small end}}  \rho_e\right)\rho(L)(1-\rho_e)  -
    k_{\mbox{\small  off}}^{\mbox{\small end}} \rho_e\right].
\label{eq:rhol}
\end{eqnarray}
We then numerically integrated eqn.~\eqref{eq:rhol} with $L_0 = 2
\times 10^5$ dimers and determined the crossover length $d$ where the
depolymerization rate decreases by 20\% below the steady-state rate.
These results are the solid curve in fig.~\ref{shorten}D. The
predictions of the mean-field model agree well with the simulations.

For these results, we find that the crossover length $d$ is 10 $\mu$m
or longer for bulk motor concentrations of 10 nM or lower. However,
for high bulk motor concentration of 50 nM or more, the crossover
length decreases to 4 $\mu$m or less. This suggests that
length-dependent depolymerization is only prominent for sufficiently
low motor concentration.  This difference may partially explain why
Varga et al.~observed length-dependent depolymerization but Gupta et
al.~did not. The Varga experiments used lower bulk motor
concentrations\cite{VargaNCB2006,GuptaNCB2006}.

\begin{figure}[t] 
    \centering 
  \includegraphics[width=10 cm]{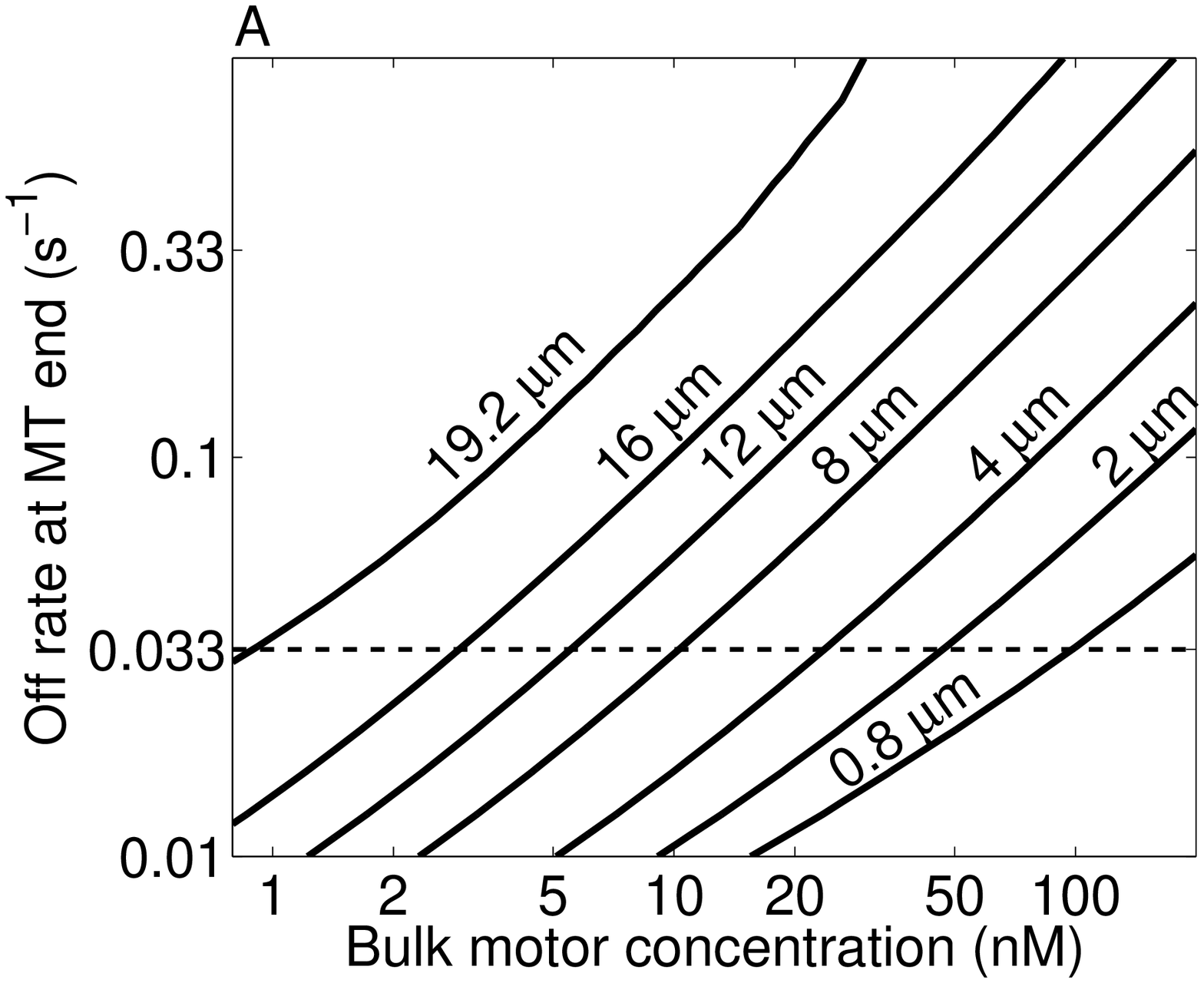}
  \includegraphics[width=10 cm]{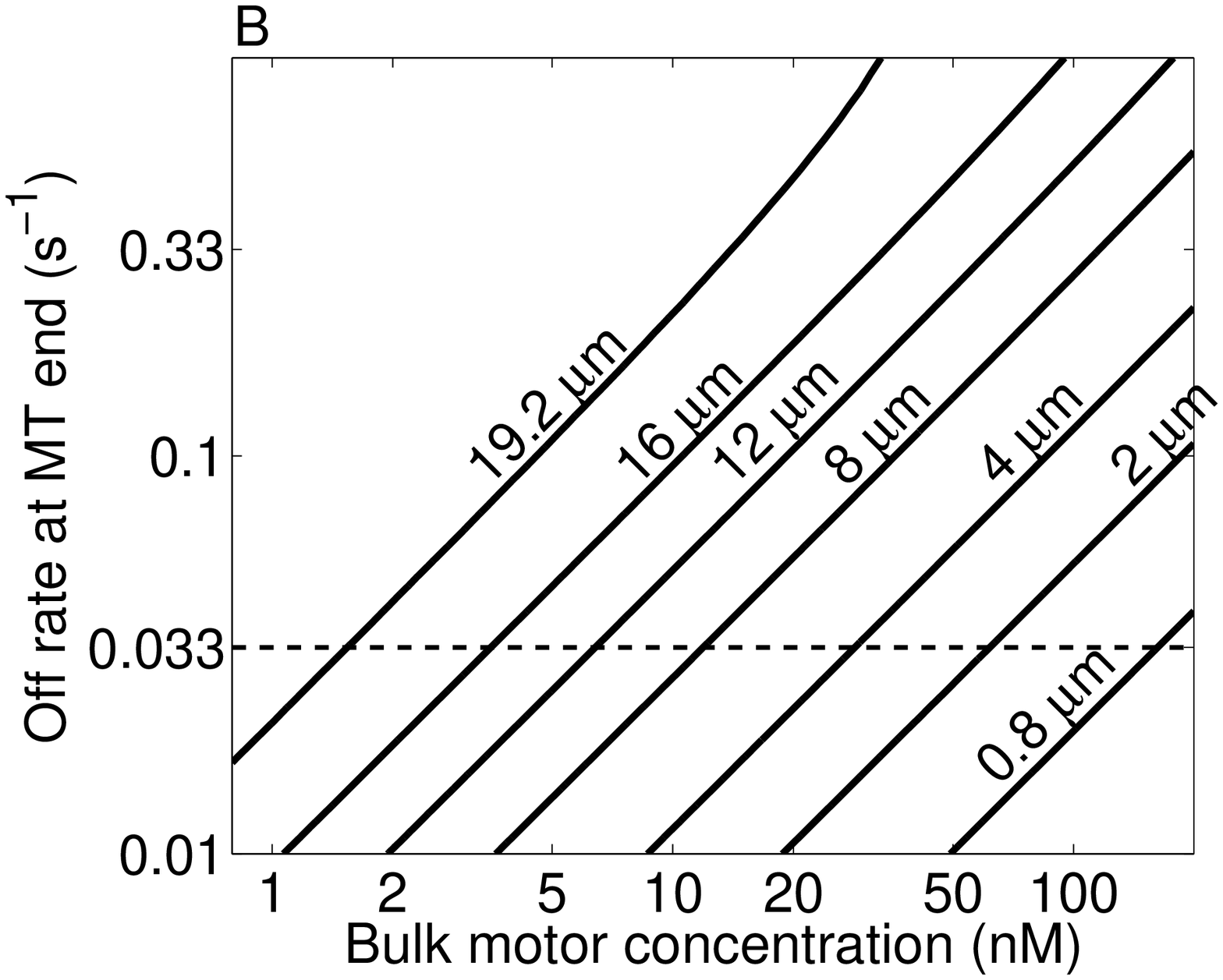}
  \caption {}
  \label{phasediagram} 
 \end{figure}

To further understand the behavior of length-dependent
depolymerization, we examined how the crossover length $d$ changes
when two key parameters are varied. We changed the bulk motor
concentration and the motor off rate at the MT end and determined $d$.
We studied this crossover using the mean-field model in two limits:
first, we numerically integrated eqn.~\eqref{eq:rhol} to determine the
crossover length in the full mean-field model. Second, we determined
the crossover length in the quasi-static model, which assumes that the
motor occupancy at the MT end is instantaneously determined by the
solution to the steady-state eqn.~\eqref{eq:rhoeqs}. This solution is
valid if the motor occupancy at the end changes quickly compared to
other time scales in the problem. The results are shown in
fig.~\ref{phasediagram}. We note that the quasi-static approximation
gives results similar to the full mean-field model except at the
lowest motor off rates. This is intuitively reasonable, since slower
motor off rate at the MT end increases the time scale for motor
dynamics at the MT end, decreasing the validity of the quasi-static
approximation.

\subsection*{Length-dependent depolymerization controlled by initial
  conditions}

The experiments of Varga et al.~do not begin with very long MTs such as
those discussed in the previous section. As a result, we expect the
dynamics to show a lag relative to the steady depolymerization
observed when starting with long MTs (see fig.~\ref{equil-trans}).  To
address conditions relevant to experiments, in this section we
characterize the effects observed in this transient regime.

For the mean-field calculations presented here, we numerically
integrated eqn.~\eqref{eq:rhoe} using a time-dependent motor density away
from the end:
\begin{eqnarray}
  \frac{d \rho_e}{d t} &=& \left(v-\frac{d L}{d
      t}\right)\rho(L,t)(1-\rho_e)   
  - k_{\mbox{\small off}}^{\mbox{\small end}} \rho_e,  \label{eq:rhotB} \\
  \frac{d L}{d t} &=&  - k_-^{\mbox{\small end}} \rho_e.
\label{eq:lt}
\end{eqnarray}
The initial conditions are $L(t=0)=L_0$ and $\rho_e(t=0)=0$.  The time
dependence of the density away from the end, $\rho(L,t)$ is
described by
\begin{equation}
\rho(x,t) = \rho_0 ( 1- e^{-x/\lambda}) (1-e^{-t/\tau}),
  \label{eq:rhotx}
\end{equation}
with $\tau = 1/( k_{\mbox{\small off}} + k_{\mbox{\small on}}c)$ and
$\lambda = v \tau$ (as discussed above; see eqns.~\eqref{eq:rhotau},
\eqref{eq:tau}, and fig.~\ref{SSdensity}).  This form of the density
is an approximation that assumes the density approaches its
steady-state distribution with dynamics controlled by the slowest
time scale in eqn.~\eqref{eq:rhoo}. Note that we assume that at time
$t=0$ motors are introduced to the system, as in the Varga et
al.~experiments.

\begin{figure}[t] 
    \centering 
  \includegraphics[width=10 cm]{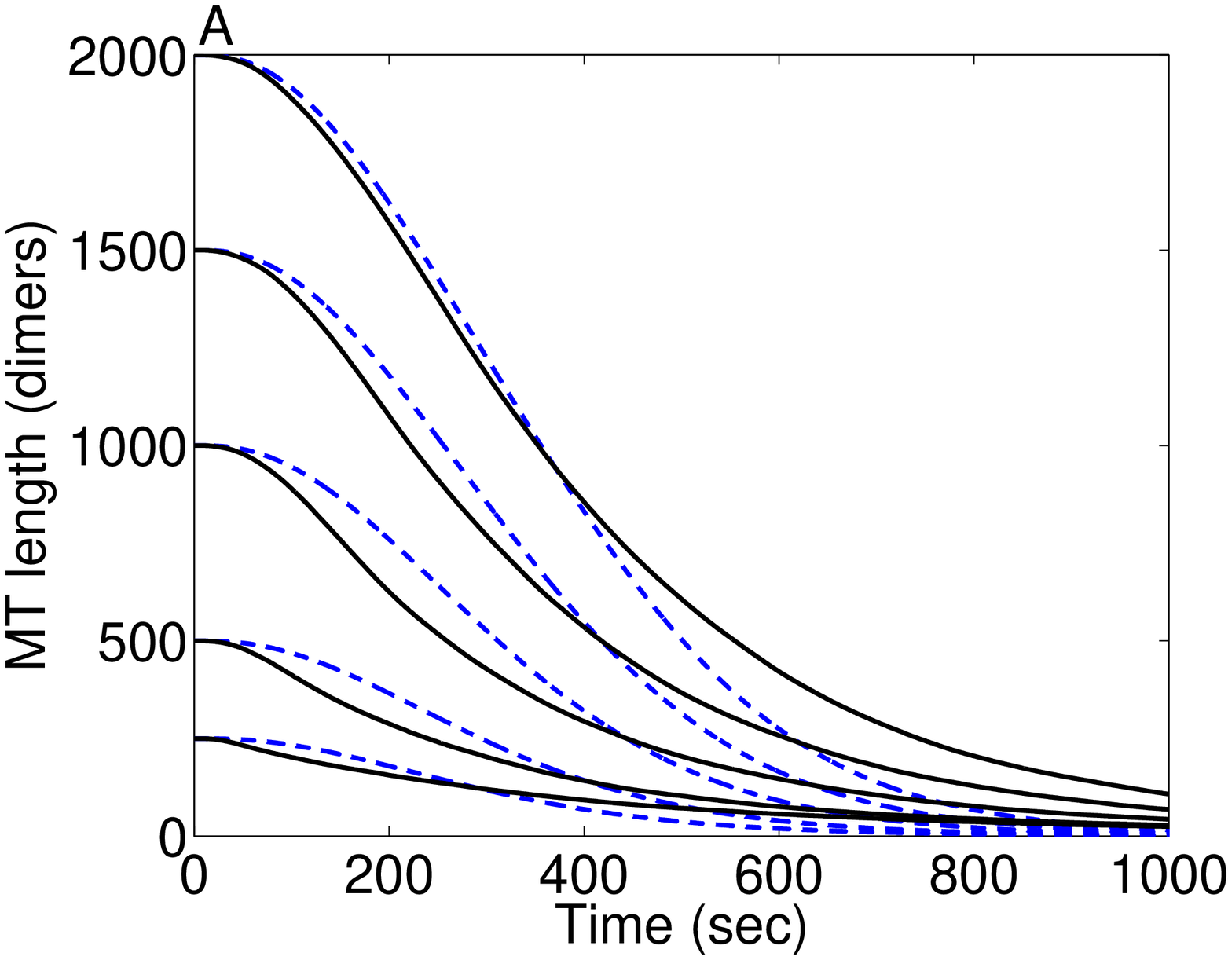}
  \includegraphics[width=10 cm]{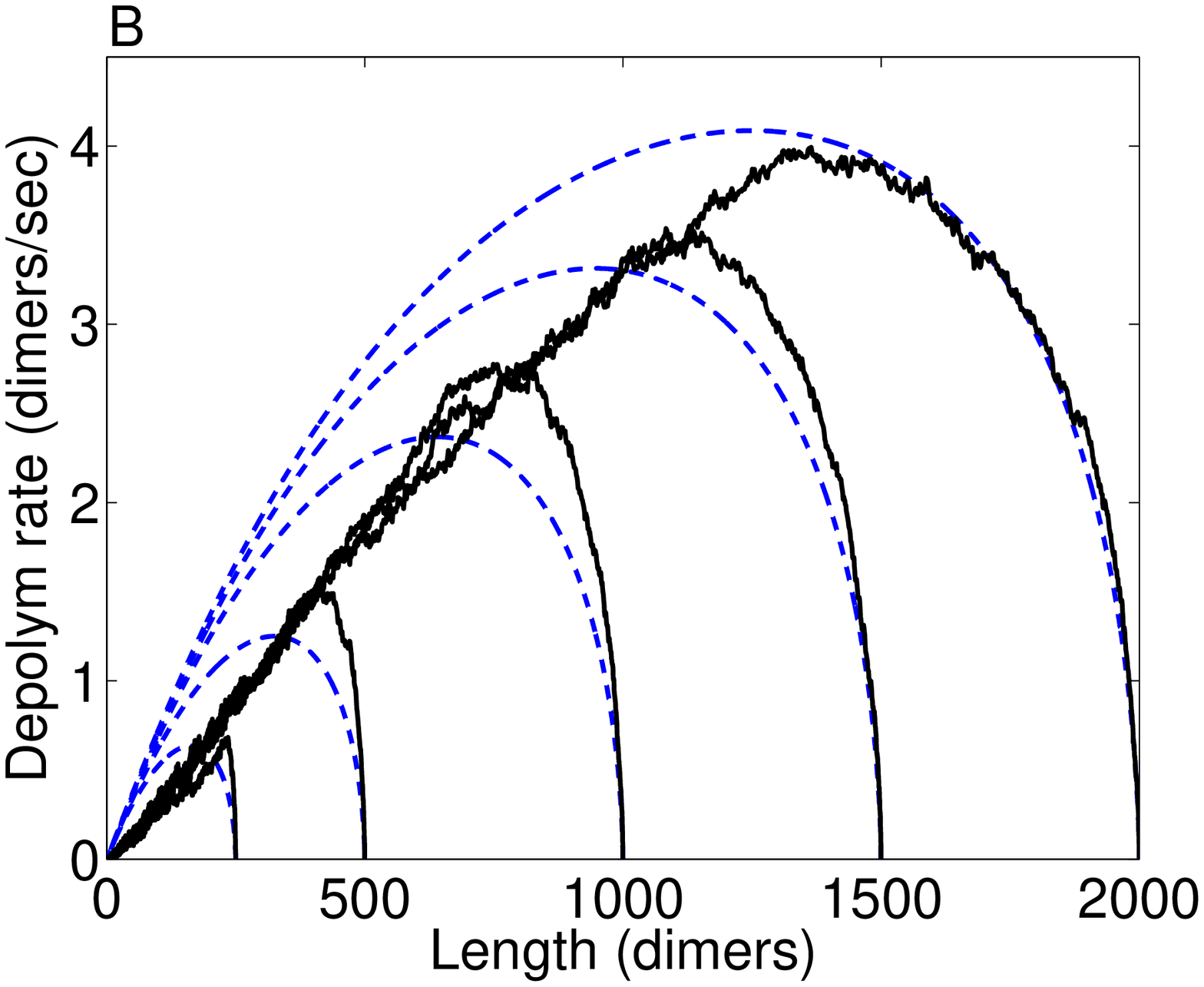}
  \caption {}
\label{length-time}
 \end{figure}

In fig.~\ref{length-time} we illustrate the solutions to these
equations for an assumed bulk motor concentration of 5 nM. The
dynamics vary with the initial length of the MT assumed; the maximum
depolymerization rate varies by almost a factor of 10 with a
comparable change in $L_0$.

We compared these results to the experimental data of Varga et al.
by considering
varying initial MT lengths and bulk motor concentrations. For each
curve, we determined the maximum depolymerization rate (the peak of
the curve in fig.~\ref{length-time}B). The results are shown in
fig.~\ref{rate-length}, where we show how the maximum depolymerization
rate varies with $L_0$ and bulk motor concentration. The slope of
depolymerization rate versus initial MT length is shown as well, with
the results of Varga et al.~shown for comparison. The results are
reasonably similar---the model predictions are within a factor of 2 of
the experimental results.

The model predictions show a linear dependence of the slope of
depolymerization rate on initial MT length, while the Varga et
al.~data suggest a non-linear dependence of the slope on bulk motor
concentration (fig.~\ref{rate-length}). This nonlinearity could result
from motor cooperativity, which would lead to a nonlinear dependence
of the depolymerization rate on the motor density at the MT end. This
would be an interesting direction for future experiments to explore.

\section*{Conclusion}

We have developed a theory of MT depolymerization by kinesin 8 and
demonstrated agreement between our theory and currently available
experiments. The model incorporates biased motor motion toward MT
plus ends, motor-catalyzed depolymerization of plus ends, motor
binding and unbinding, and motor crowding effects.  Our theory
quantitatively reproduces the experiments of Varga et al.~using
purified Kip3p, the budding yeast kinesin 8.

In developing a theory of kinesin-8 motors, we addressed a limit which
has not been considered in previous theoretical work
\cite{parmeggiani04,nowak07,KleinPRL2005,govindan08}.  Experiments
have revealed that clumps of kinesin 8 form on the MT plus end and
change significantly during MT
depolymerization\cite{VargaNCB2006,GuptaNCB2006}.  This observation
suggests that MT shortening and motor density changes occur on similar
time scales.  Therefore a steady-state mathematical analysis is likely
to be poor approximation to experiments. We have therefore developed
and analyzed a time-dependent equation for the MT end that couples MT
depolymerization, motor arrival and dissociation, and motor crowding
effects. The results of the equation for MT end dynamics agree well
with full Monte Carlo simulations of the model.

Despite the good agreement between our theoretical results and
experiments, there are several uncertainties in our model.  We have
estimated the motor binding rate constant, which has not been
precisely quantified in experiments. The most significant effect of
this uncertainty is that it leads to uncertainty in the
depolymerization processivity, as illustrated in fig.~\ref{variance}.
In addition, changing the on rate constant would alter the dependence
of the crossover length $d$ on motor concentration, changing the
predictions of fig.~\ref{phasediagram}. We also assumed that when a
motor catalyzes dimer removal, it falls off the MT if another motor is
bound directly behind it. This assumption is supported by \textit{in
  vivo} experiments with fluorescently-labeled Kip3p, which show that
a clump of motors accumulates at the MT plus end when the MT is
growing, but the clump shrinks during MT shortening
\cite{GuptaNCB2006,VargaNCB2006}.  Finally, we did not consider
possible effects of motor cooperativity in depolymerization;
cooperativity is a possible explanation for the apparent nonlinearity
of the experimental data shown in fig.~\ref{rate-length}. Our best-fit
parameters did lead to quantitative agreement between the theory and
the experiments of Varga et al. Errors in our determination of the
parameters would lead to changes in the quantitative predictions of
the theory, but we verified that the qualitative predictions are
insensitive to the exact parameter values chosen.

We observed length-dependent depolymerization of MTs in the model, as
seen experimentally by Varga et al.~\cite{VargaNCB2006} and
theoretically by Govindan et al.~\cite{govindan08}. However, this
phenomenon occurs only below a ``crossover'' length $d$; MTs much
longer than $d$ will depolymerize at a constant rate. The crossover
length is controlled by key parameters of the model, particularly the
bulk motor concentration, the motor translocation processivity, and
the depolymerization processivity.  In particular, using the best-fit
experimental parameters we predict that the crossover length will
decrease from 16 $\mu$m to 4 $\mu$m if the bulk motor concentration is
increased from 5 to 50 nM. This strong concentration dependence may
explain the observation of length-dependent depolymerization
\textit{in vitro} by Varga et al.~\cite{VargaNCB2006} and not by Gupta
et al.~\cite{GuptaNCB2006}.  (Note that the experiments differed in
other ways, such as the observed motor velocity.)

Length-dependent depolymerization is not specific to motors with
biased motility \cite{govindan08}. For example, length-dependent
depolymerization could occur for the kinesin-13 MCAK, which diffuses
on MTs.  The crossover to length-dependent depolymerization is
sensitive to motor processivity. As a result, we expect
length-dependent depolymerization will be much more important in the
kinesin-8 family of proteins because they have a much higher
processivity than do kinesin 13s ($\sim 12\ \mu$m for Kip3p vs $\sim 1
\ \mu$m for MCAK) \cite{GuptaNCB2006}.

The fact that the crossover length is so parameter dependent
highlights the need for care in interpreting experimental results.  In
cells, the bulk motor concentration and other parameters may be
different from the \textit{in vitro} values. If the key features of
our model do apply in the more complicated cellular environment, we
can make some speculative predictions. In particular, an
overexpression experiment, which increases motor concentration, would
lead to both shorter MTs and to a decrease in the crossover length
$d$, and therefore a decrease in the range of lengths over which there
is length-dependent depolymerization.

Our results are most consistent with processive depolymerization by
kinesin-8 motors, suggesting that multiple tubulin dimers can be
removed by a single motor. In particular, the experimental data can be
fit to a non-processive model only if (i) the motor on rate is a
factor of 20 higher than the one estimated from experiments, and (ii)
GMPCPP MTs are intrinsically unstable, so removal of one dimer from
one row of tubulin subunits leads to the removal of all 13 dimers
around the MT.  We note that processive depolymerization has been
observed for the kinesin-13 motor MCAK \cite{HeleniusNature2006}, so
kinesin 13s and 8s may share some features in their depolymerization
activity.

We have shown that processive depolymerization tends to increase the
fluctuations of MT length about its average (fig.~\ref{variance}).
Therefore the roughness of the MT length versus time during
depolymerization could be used to quantify the depolymerization
processivity. While we emphasize that our theory may not be directly
relevant in cells, our results on MT length fluctuations do have
interesting implications for some recent experiments. Although kinesin
8s serve, on average, to decrease MT length, the large fluctuations
that result from processive depolymerization would introduce a
significant variance about this average behavior. In addition, higher
protein concentrations should decrease these fluctuations
(fig.~\ref{shorten}C). If this were true in cells, it could explain
two recent and puzzling observations on kinesin 8 \textit{in vivo}.
Stumpff et al.~found that kinesin 8 over-expression decreased the
amplitude of metaphase chromosome oscillations, while reduction in
kinesin-8 concentration by RNAi increased it \cite{stumpff08}.
Unsworth et al.~found that deleting the kinesin-8 genes (completely
 eliminating the protein) from fission yeast decreased MT length
fluctuations \cite{unsworth08}, as would be expected if physiological
levels of kinesin 8 contribute significantly to length fluctuations.
Although our simple mathematical model does not describe the full
complexity of the mitotic spindle, it does highlight the potential
role of motor-induced fluctuations in spindle behavior.

\section*{Acknowledgements}

This work was supported in part by an NIH Training Grant (GM65103,
LEH) and the Alfred P. Sloan Foundation (MDB).


\newpage
\subsubsection*{Table 1.}
\begin{center}
  \begin{tabular*}{1.0\textwidth}{@{\extracolsep{\fill}} 
| m{2.2in} l m{1.6in}| }
    \hline Quantity & Symbol  & Typical value(s)\\ \hline
    \hline

    Position along MT & $x$ & 0 -- 2500 dimers \\
    \hline

    MT length & $L$  & 0 -- 2500 dimers \\
    \hline

    Motor velocity & $v$  & 7.5 dimers s$^{-1}$\\
    \hline 

    Motor on rate constant & $k_{\mbox{\small on}}$  & $ 3 \times 10^{-6}$
    site$^{-1}$ nM$^{-1}$ 
    s$^{-1}$ \\ \hline

    Motor off rate & $k_{\mbox{\small
        off}}$  & $ 5 \times 10^{-3}$ s$^{-1}$ \\
    \hline

    Equilibrium constant for motor binding to MT & $K =
    k_{\mbox{\small off}}/k_{\mbox{\small on}}$  & 1.67 $\mu$M
    \\ \hline

    Motor run length & $\ell = v/k_{\mbox{\small off}}$  &
    1500 dimers\\ \hline

    Motor off rate at MT end& $k_{\mbox{\small off}}^{\mbox{\small
        end}}$ & $ 3.3 \times 10^{-2}$ s$^{-1}$ \\
    \hline

    Rate of motor-catalyzed MT depolymerization &
    $k_-^{\mbox{\small end}}$  & 13 dimers s$^{-1}$ \\
    \hline

    Motor depolymerization processivity (upper bound) & $a = k_-^{\mbox{\small
        end}}/k_{\mbox{\small off}}^{\mbox{\small end}}$  & 1
    -- 400 dimers \\ \hline

    Bulk motor concentration & $c$  & 1 -- 200 nM \\ \hline

    Motor occupancy per tubulin dimer & $\rho(x)$  & 0
    -- 1 \\ \hline

    Steady-state motor occupancy away from MT ends & $\rho_0 =
    k_{\mbox{\small on}}c/(k_{\mbox{\small on}}c+k_{\mbox{\small
        off}})$  & $ 6 \times 10^{-4}$ -- $ 6 \times
    10^{-2}$ \\ \hline

    Motor occupancy at MT plus end & $\rho_e$  & 0 -- 1 \\ \hline

    Motor occupancy boundary length & $\lambda=v/(k_{\mbox{\small
        on}}c+k_{\mbox{\small off}})$  & 1400 -- 1500 dimers \\
    \hline

    Time scale of approach to steady-state motor occupancy &$\tau =
    1/(k_{\mbox{\small on}}c+k_{\mbox{\small off}})$  & 185
    -- 200 seconds \\
    \hline

    Length of crossover to length-dependent depolymerization 
    & $d$  & 0 -- 2400 dimers \\ \hline

    Bulk tubulin concentration & $c_t$  & 10 -- 100 nM \\ \hline

    MT polymerization rate constant & $\alpha$ & 5.4 dimers
    $\mu$M$^{-1}$ s$^{-1}$ \\ \hline 

    MT depolymerization rate & $\beta$  & 0.1 dimers
    s$^{-1}$ \\ \hline
  \end{tabular*}
\end{center}

\newpage
\section*{Figure Legends}

\subsubsection*{Figure~\ref{model}.}
Model of kinesin-8 motor protein's interaction with a MT protofilament
showing the key rates.  Rates in black affect only the motor, while
those in red affect the MT plus end.  The plus end of the MT
protofilament is indicated by a thick vertical (red) line, and the
dimers are indicated by (blue) boxes.  Note that depolymerization (at
rate $k_{-}^{\mbox{\small end}}$) affects both the motor and the MT
plus end.  (A) A motor binds to a dimer of the MT with on rate
$k_{\mbox{\small on}} c$ and unbinds with rate $k_{\mbox{\small
    off}}$.  The motor steps forward at rate $v$; backward motion is
not considered due to the biased motion of kinesin 8s.  MT dynamics
are represented by allowing dimers to add to a MT end at rate $\alpha
c_t$ (where $c_t$ is the bulk concentration of tubulin dimers) and
dissociate at rate $\beta$.  (B-D) Depolymerization models.  (B) If
the motor depolymerizes processively, it removes a MT dimer as it
steps backward (with rate $k_{-}^{\mbox{\small end}}$), thereby
shortening the MT.  (C) If the dimer behind the MT end is occupied,
the motor falls off the MT in either model. (D) In the nonprocessive
depolymerization model, the motor removes a single tubulin dimer and
falls off the MT.  (E) Lateral interactions help stabilize MTs.  We
incorporated this into our model by allowing the depolymerization rate
to depend on the number of neighboring protofilaments (i.e., 0, 1, or
2).  In this case, the rate at which a terminal tubulin dimer unbinds
from a protofilament is given by $\beta_0$ if the dimer has no lateral
neighbors, $\beta_1$ if the dimer has 1 lateral neighbor, and
$\beta_2$ if the dimer has 2 lateral neighbors.

\subsubsection*{Figure~\ref{longtails}.}
The slow polymerization of GMPCPP MTs in the presence of free tubulin
accounts for the long-time tails observed in Varga \textit{et
  al.}~\cite{VargaNCB2006}.  In that work, Kip3p was unable to fully
depolymerize the MTs over the course of a single experiment.  In our
simulations, adding previously measured MT polymerization and
depolymerization rates for GMPCPP stabilized MTs \cite{pmid1421572}
reproduced the observed behavior. The solid traces were made assuming
that the only MT dynamics were those caused by the motors, while the
dashed traces were made including intrinsic MT polymerization and
depolymerization.

\subsubsection*{Figure~\ref{variance}.}
Effects of depolymerization processivity on MT length fluctuations.
We chose model parameters which led to similar overall behavior of MT
length versus time (A,C) but had different motor-induced
depolymerization rates and motor on rates. All simulations of model P
used the experimentally derived value $k_{\mbox{\small
    off}}^{\mbox{\small end}} = 3.3 \times 10^{-2}$ s$^{-1}$. The
maximum processivity (the maximum numbers of dimers removed per motor)
is $a=k_-^{\mbox{\small end}} / k_{\mbox{\small off}}^{\mbox{\small
    end}}$.  (A) We first considered completely stable MTs, those with
$\beta_{0,1,2}=0$, and varied $k_{\mbox{\small on}}$ and
$k_-^{\mbox{\small end}}$ to obtain curves which all have a similar
average shape.  The trace with $a=394$ uses $c=1$ nM and the best-fit
parameters found when comparing to experiments: $k_{\mbox{\small on}}
= 3 \times 10^{-6}$ site$^{-1}$ nM$^{-1}$ s$^{-1}$ and
$k_-^{\mbox{\small end}} = 13$ dimers s$^{-1}$.  The curve with
$a=197$ has the on rate constant doubled to $6 \times 10^{-6}$
site$^{-1}$ nM$^{-1}$ s$^{-1}$ and the maximum depolymerization rate
halved to $6.5$ dimers s$^{-1}$.  The curve with $a=52$ has the on
rate constant increased by a factor of 8 to $k_{\mbox{\small on}} = 24
\times 10^{-6}$ site$^{-1}$ nM$^{-1}$ s$^{-1}$ and $k_-^{\mbox{\small
    end}} = 1.7$ dimers s$^{-1}$.  Curves in the main panel show
averages of 500 simulations, each of a MT with 13 independent
protofilaments.  Curves in the inset panel show results of individual
simulations; the roughness of the MT length versus time behavior
decreases as $a$ is decreased.  (B) The standard deviation of MT
length as a function of time for the simulations shown in (A).
Simulations with higher $a$ show larger standard deviation.  (C)
Allowing protofilaments without two neighbors to spontaneously
depolymerize (P, $k_{\mbox{\small on}}$, high $\beta$) gives a similar
average curve to a simulation with fully stable MTs, but
$k_{\mbox{\small on}}$ increased by a factor of 13.  A similar average
curve can also be obtained in the non-processive case, but only if
protofilaments without two neighbors spontaneously depolymerize and
$k_{\mbox{\small on}}$ is increased by a factor of 300 (NP, $300\times
k_{\mbox{\small on}}$, high $\beta$).  (D) The standard deviation of
MT length for the simulations shown in (C).

\subsubsection*{Figure~\ref{SSdensity}.}
(A) Examples of motor occupancy profiles. For this figure, we assumed
a much higher on rate than found in experiments to make occupancy
changes visible. The reference parameter set has $k_{\mbox{\small on}}
c = 0.002 $ dimer$^{-1}$ s$^{-1}$ (which would correspond to a bulk
motor concentration of $667$ nM at the typical on rate constant of
$k_{\mbox{\small on}} = 3 \times 10^{-6} $ dimer$^{-1}$ s$^{-1}$
nM$^{-1}$), $k_{\mbox{\small off}} = 0.005 $ s$^{-1}$, and
$k_{\mbox{\small off}}^{\mbox{\small end}} = 0.02$ s$^{-1}$ (black
solid line).  Therefore $\ell = 1071$ dimers (black vertical line).
The curve with decreased $\ell$ (blue dotted line) has
$k_{\mbox{\small on}} c $ and $k_{\mbox{\small off}}$ both doubled to
halve $\ell$ to 536 dimers (blue vertical dashed-dotted line) while
keeping $K_D$ unchanged.  For these parameters, the mean-field
expression for the occupancy from eqn.~\eqref{eq:rho} is the blue
dashed/dotted line. The curve with decreased $K_D$ has
$k_{\mbox{\small on}} c $ halved (red dashed line).  (B) Average motor
density as a function of time for the lower $\ell$ parameter set at
three positions along the MT: $\ell/2$ (black solid line), $3\ell/2$
(red solid line) with the mean-field analytic expression of
eqn.~\eqref{eq:rhotau} superimposed (black dotted line), and the MT
plus end (blue dashed-dotted line).  All plots are averages of 500
simulated MTs.

\subsubsection*{Figure~\ref{shortenss}.}
Steady-state motor occupancy and depolymerization rate of long MTs.
Left axis shows steady-state motor occupancy at the MT plus end (solid
blue line, black circles) or away from the end (red dashed-dotted
line) as a function of bulk motor concentration. The mean-field model
(solid blue line) is the prediction of equation \eqref{eq:rhoess} for
the steady-state occupancy of the MT plus end.  Right axis shows the
resulting steady depolymerization rate in the mean-field model
(solid blue line) and simulations (black circles).  The simulation
results were determined from the simulations shown in
fig.~\ref{shorten}.

\subsubsection*{Figure~\ref{timeconst}.}
Approach to steady-state end occupancy in the mean-field model. (A)
End occupancy, beginning from $\rho_e(t=0)=0$, for representative
values of the bulk motor concentration. (B) Values of the time
constants $r_{\pm}^{-1}$ as a function of bulk motor concentration.

\subsubsection*{Figure~\ref{equil-trans}.}
Steady-state versus transient depolymerization. The transient
condition (when motors are added at $t=0$, red dashed line) gives
dynamics with an initial lag when compared to the condition with
pre-equilibrated motors (blue solid line).  Results shown are from
simulations (average of 500 runs) with bulk motor concentration of 5.5
nM.  (A) MT length as a function of time.  (B) Depolymerization rate
as a function of MT length.

\subsubsection*{Figure~\ref{shorten}.}
Depolymerization of long MTs.  Each curve is the average of 500
independent simulations. Each simulation was started from a
pre-equilibrated MT: the simulation was run for 1000 sec with no
filament depolymerization, to allow the motor density on the MT to
reach steady state. (A) Length versus time.  (B) Depolymerization rate
versus MT length. Black squares indicate the crossover to
length-dependent depolymerization. (C) Standard deviation of MT length
versus MT length.  (D) Length of crossover to length-dependent
depolymerization in the simulations (red circles) and the mean-field
model (blue solid line).

\subsubsection*{Figure~\ref{phasediagram}.}
Length of crossover to length-dependent depolymerization, as a
function of the bulk motor concentration and the motor off rate at the
MT end. The horizontal dashed line is the motor off rate at the MT end
estimated from the experiments of Varga et al. (A) Mean-field model.
(B) Quasi-static approximation, where the motor occupancy at the MT
end is assumed to change quickly compared to other dynamics in the
problem.

\subsubsection*{Figure~\ref{length-time}.}
Dynamics under conditions of transient shortening depend strongly on
the initial MT length. Black solid lines: simulations (average of 500
runs). Blue dashed lines: mean-field model. At $t=0$,
motors are introduced to the system. The bulk motor concentration was
5 nM.  (A) MT length as a function of time.  (B) Depolymerization rate
as a function of MT length. 

\begin{figure}[t] 
    \centering 
  \includegraphics[width=8.5 cm]{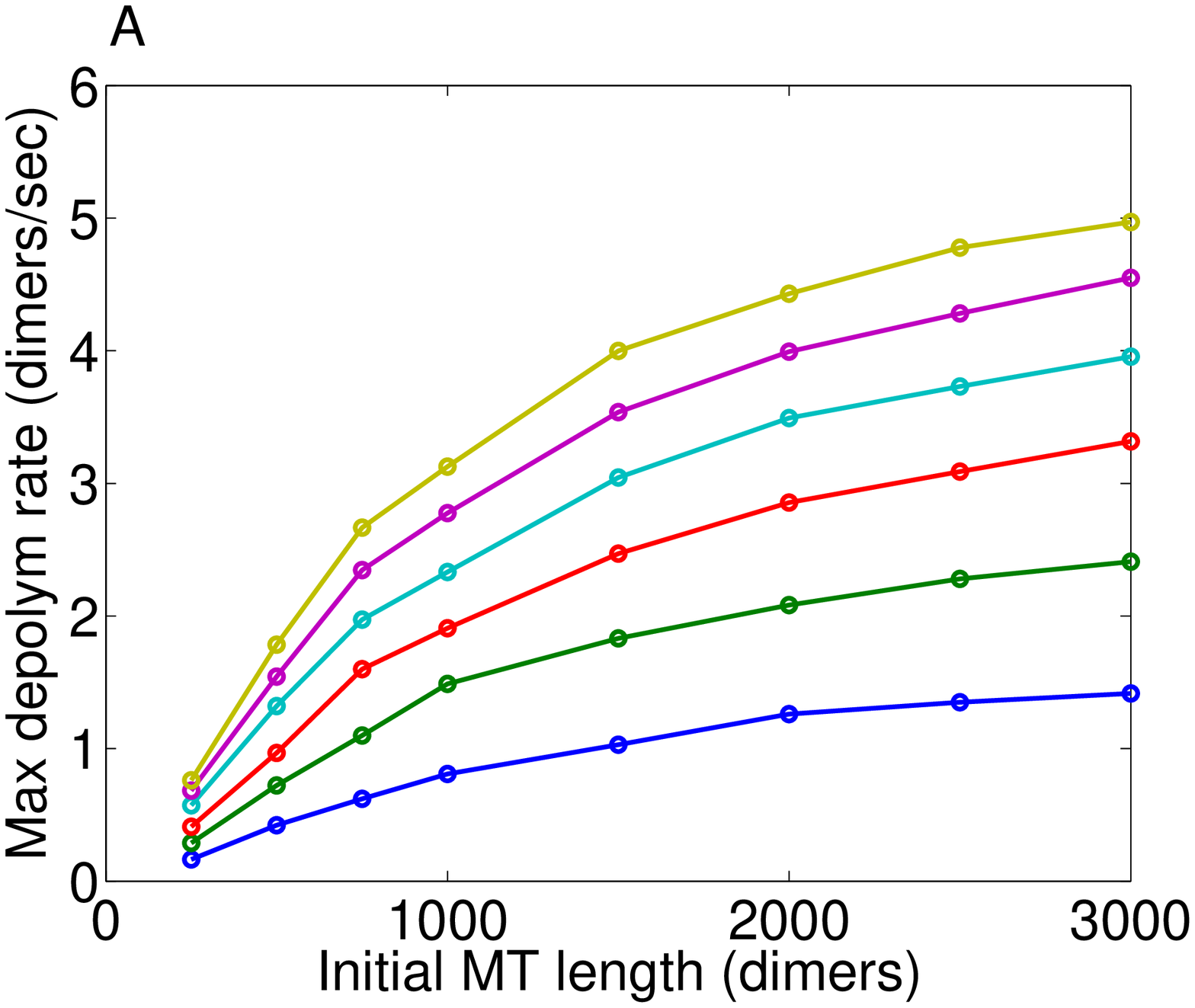}
  \includegraphics[width=8.5 cm]{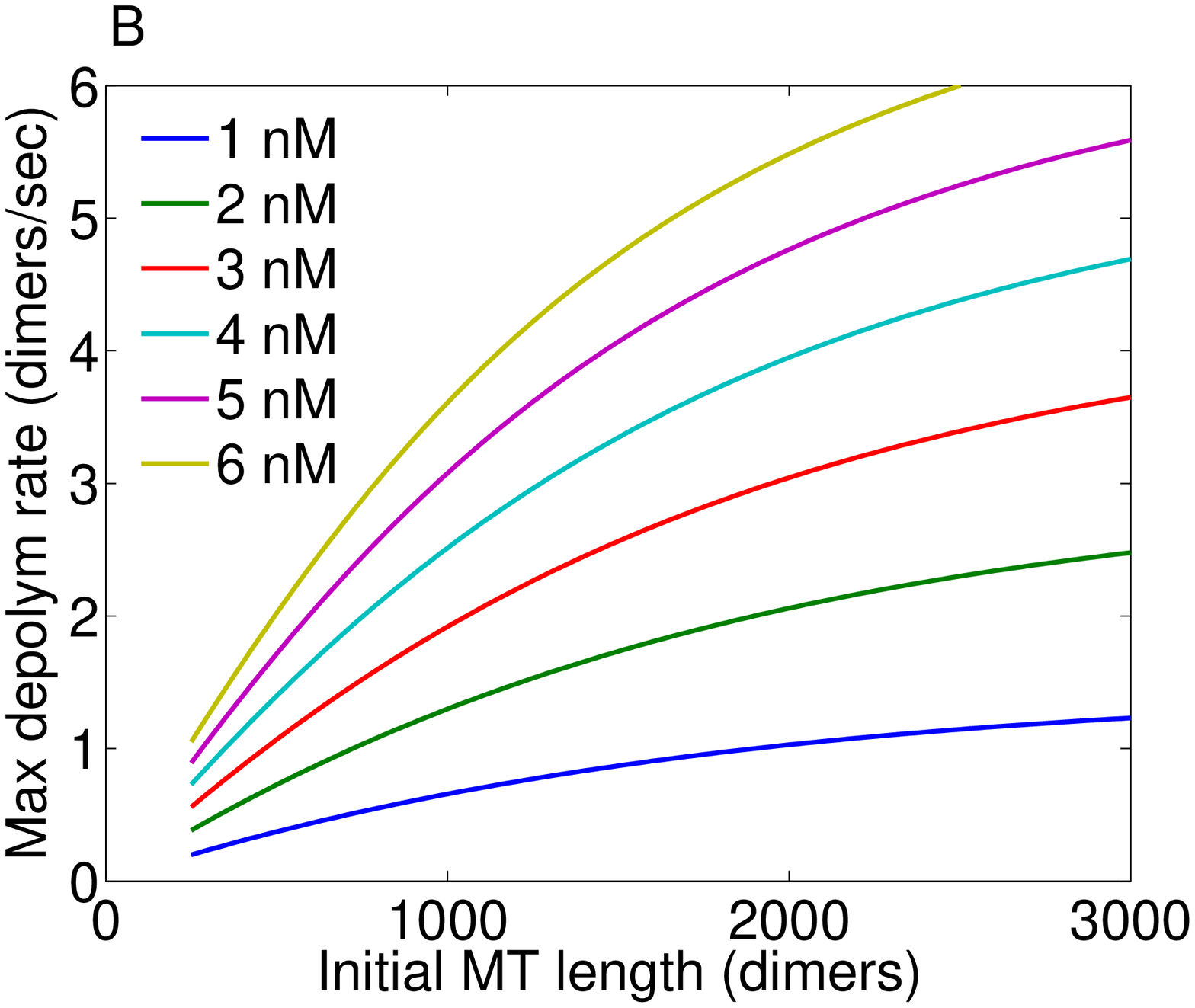}
  \includegraphics[width=8.5 cm]{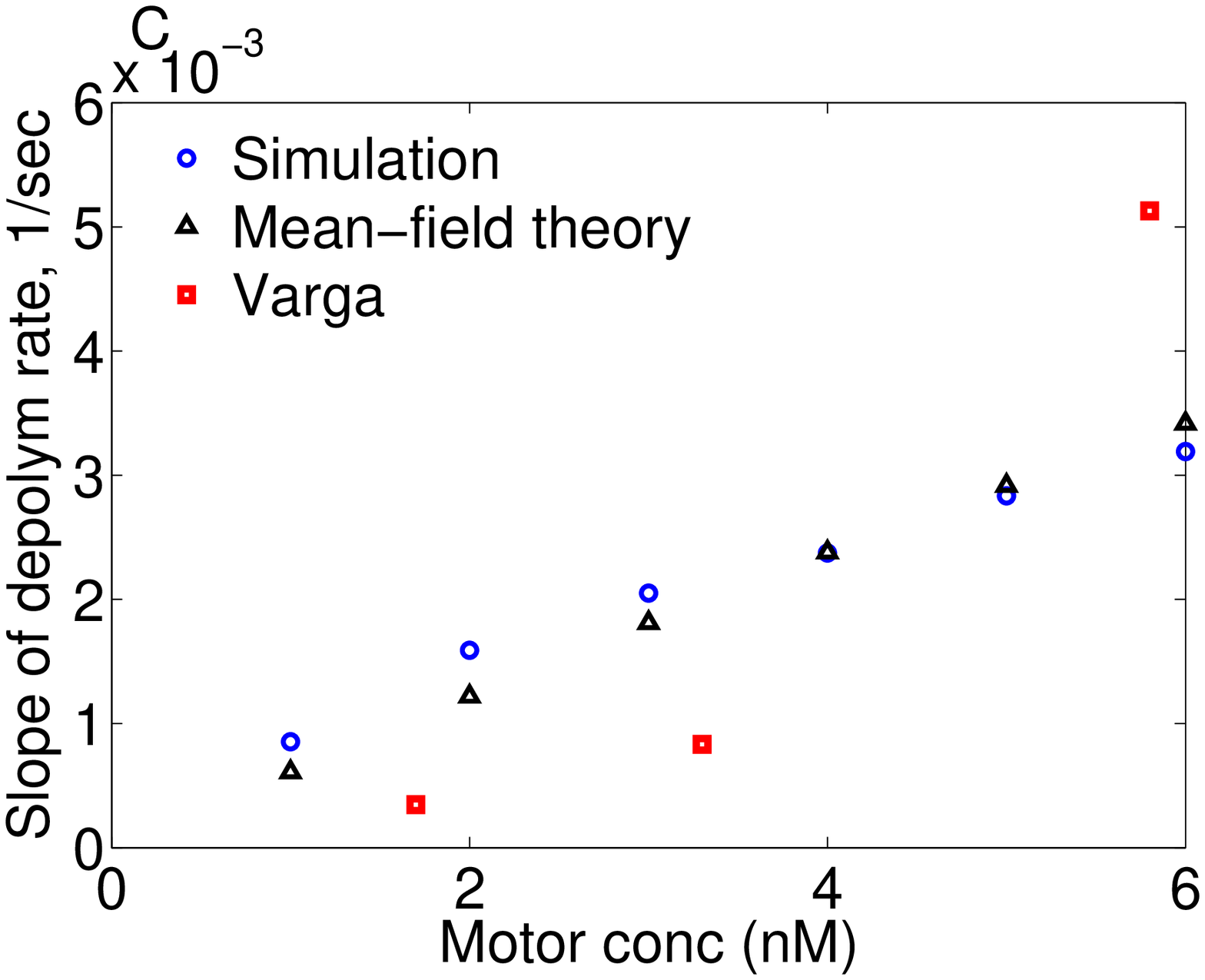}
  \caption {}
\label{rate-length}
 \end{figure}

\subsubsection*{Figure~\ref{rate-length}.}
Variation of the maximum depolymerization rate with initial MT length,
showing predictions of (A) the simulations and (B) the mean-field
model under conditions of transient shortening.  (C) To compare to the
results of Varga et al., we fit a straight line to the first 1000
dimers of each curve in (A-B), and determined the slope as a function
of bulk motor concentration.  The data of Varga et
al.~\cite{VargaNCB2006} are shown for comparison.

\clearpage

\clearpage

\clearpage

\clearpage

\clearpage

\clearpage

\clearpage

\clearpage

\clearpage

\clearpage

\clearpage

\end{document}